\documentclass[lettersize,journal]{IEEEtran}
\usepackage{amsmath,amsfonts}
\usepackage[ruled,vlined,linesnumbered]{algorithm2e}
\usepackage{array}
\usepackage{subcaption}
\usepackage{xcolor}
\usepackage{textcomp}
\usepackage{stfloats}
\usepackage{url}
\usepackage{verbatim}
\usepackage{graphicx}
\usepackage{cite}
\usepackage{multirow}
\usepackage{makecell}
\usepackage{booktabs}   

\begin{document}

\title{
\textit{BEACON}: Benefit-Aware Early-Exit for Automatic Modulation Classification via Recoverability Prediction
}

\author{
Zheng~Liu,
Hatem~Abou-Zeid,~\IEEEmembership{Member,~IEEE,}
and Huaqing~Wu,~\IEEEmembership{Member,~IEEE}
\thanks{ \scriptsize Zheng Liu, Hatem~Abou-Zeid, and Huaqing Wu are with the Department of Electrical and Software Engineering, University of Calgary, Calgary, AB T2N 1N4, Canada (e-mail: \{zheng.liu1, hatem.abouzeid, huaqing.wu1\}@ucalgary.ca). Copyright (c) 2026 IEEE. Personal use of this material is permitted. However, permission to use this material for any other purposes must be obtained from the IEEE by sending a request to pubs-permissions@ieee.org. 

This is the accepted version of the article published in IEEE Internet of Things Journal, DOI: 10.1109/JIOT.2026.3710690}
}



\IEEEaftertitletext{\vspace{-1\baselineskip}}
\maketitle

\begin{abstract}
Convolutional neural networks (CNNs) have emerged as a powerful tool for automatic modulation classification (AMC) by directly extracting discriminative features from raw in-phase and quadrature (I/Q) signals. However, deploying CNN-based AMC models on IoT devices remains challenging because of limited computational resources, energy constraints, and real-time processing requirements. 
Early-exit (EE) strategies alleviate this burden by allowing qualified samples to terminate inference at an EE branch. However, our empirical analysis  reveals a critical limitation of existing confidence-based EE strategies: they predominantly select samples whose early and final predictions are  correct and consistent, while failing to capture whether deeper inference can provide a tangible accuracy gain.  
To address this limitation, we propose \textit{BEACON}, a \underline{B}enefit-aware \underline{E}arly-exit framework for \underline{A}M\underline{C} via rec\underline{O}verability predictio\underline{N}. \textit{BEACON} introduces
a benefit-aware EE criterion that explicitly predicts \emph{recoverable errors}, defined as instances where the final-exit branch corrects an initial early-branch misclassification. Using only short-branch observables, we design a lightweight benefit-aware predictor (LBAP) to implement this criterion, estimating the likelihood of such recoverable cases and triggering deeper inference only when an accuracy gain is expected. Extensive experiments on ResNet-18-based AMC models demonstrate that the proposed approach consistently outperforms state-of-the-art baselines, achieving a superior accuracy–computation trade-off across diverse EE threshold settings and signal-to-noise ratio regimes. These findings validate the effectiveness of the benefit-aware criterion and its practicality for energy-efficient on-device AMC under stringent resource constraints.

\end{abstract}


\section{Introduction}

Automatic modulation classification (AMC) is a fundamental physical-layer task in wireless communication systems, aiming to identify the modulation type of received signals under varying signal-to-noise ratio (SNR) conditions. Accurate AMC enables a wide range of communication functions, including spectrum awareness, adaptive transmission, interference management, and intelligent signal processing \cite{dobre2007survey, zheng2025recent}. 

With the rapid growth of data-driven communication systems, deep learning (DL) techniques have been widely applied to AMC to enhance classification robustness under complex channel conditions. In particular, convolutional neural networks (CNNs) have demonstrated strong capability in extracting discriminative features from raw in-phase and quadrature (I/Q) signal samples. Recent CNN-based AMC models have achieved remarkable classification accuracy across diverse SNR regimes and modulation schemes, significantly outperforming traditional feature-based methods \cite{zhang2020cnn, wang2021cnn,liu2022amc, chen2023amc}.

In IoT communication systems, AMC is often part of a closed-loop communication process, where the predicted modulation type is immediately used for subsequent operations such as signal decoding, demodulation adaptation, and spectrum access control. Excessive inference latency may therefore lead to outdated decisions or decoding failures under fast-changing wireless environments. Prior real-time AMC implementations have shown that practical RF systems may require classification latency on the order of several microseconds for cognitive radio applications \cite{tridgell2020rfsoc}. In addition, AMC is widely used in cognitive radio, spectrum monitoring, and military communication systems, where transmitting raw I/Q samples to external servers may introduce privacy and security risks \cite{alwarafy2021survey}. Therefore, efficient on-device AMC is highly desirable to reduce latency, communication overhead, and exposure of sensitive RF information. Consequently, deploying efficient and lightweight AMC models becomes a fundamental challenge, as such devices often operate under limited computational resources, tight energy budgets, and stringent real-time processing constraints \cite{duan2022distributed,li2025integration,hallaq2025tiny,liu2024user}. The high inference cost of deep neural networks can therefore hinder the practical deployment of AMC on these resource constrained devices. To address this challenge, early-exit (EE) architectures have been introduced to reduce inference complexity by allowing qualified samples to terminate inference at EE branches, thereby avoiding unnecessary computation in deeper network layers \cite{teerapittayanon2016branchynet, kaya2019overthinking}. Recent studies have applied EE mechanisms to AMC to accelerate inference while preserving classification accuracy \cite{elsayed2023earlyexit}. 

A critical component of EE  architectures is the exit criterion, which decides whether a sample should terminate early or continue to deeper layers. Existing criteria \cite{Rahmath2024EarlyExitSurvey} are primarily based on confidence-related measures, such as entropy, maximum softmax probability (MSP), or margin, which characterize the output probability distribution of the EE branch and assess whether a single class dominates the prediction. 

In this work, we first conduct an empirical analysis that reveals a fundamental inefficiency of such confidence-based criteria. Specifically, they predominantly trigger early exit when shallow-layer confidence is already high. In such instances,  the EE and final-exit (FE) predictions are typically correct and consistent.  However, these criteria overlook other scenarios where early exiting is also desirable. From a decision-theoretic perspective, invoking deeper inference is beneficial only when an EE error can be corrected by the FE prediction. In contrast, deeper inference is unnecessary when the early prediction is already correct and ineffective when both early and final predictions are incorrect.  

Motivated by these insights, we propose \textit{BEACON}, a \underline{B}enefit-aware \underline{E}arly-exit framework for \underline{A}M\underline{C} via rec\underline{O}verability predictio\underline{N}. \textit{BEACON} introduces a novel \emph{benefit-aware}  EE criterion that explicitly predicts \emph{recoverable errors}, defined as cases where the EE prediction is incorrect while the FE prediction is correct. Unlike conventional approaches, this criterion directly predicts the potential benefit of invoking deeper layers. To implement this,  we design
a lightweight benefit-aware predictor (LBAP), a compact four-layer fully connected network. The LBAP processes observable probability distribution features from the EE branch to estimate the probability that a sample belongs to the recoverable error category.

We validate the performance of \textit{BEACON} on three ResNet-18-based AMC EE models with different exit positions. 
Across diverse exit threshold settings, \textit{BEACON} consistently achieves a better accuracy-computation trade-off than representative confidence-based baselines.
Notably, \textit{BEACON} achieves up to a \textbf{24\%} improvement in overall accuracy under the same computational constraint, while baseline methods require up to \textbf{2.98$\times$} higher computation cost to reach the same accuracy.
Moreover, under a fixed FE invocation rate, \textit{BEACON} forwards a higher proportion of recoverable-error samples  to deeper layers, leading to more effective utilization of additional computation. 
To further assess robustness, we conduct  SNR-dependent evaluations under varying channel conditions, demonstrating that \textit{BEACON} maintains stable and superior performance across different SNR regimes. These results highlight \textit{BEACON}'s suitability for practical on-device AMC deployment in dynamic and resource-constrained IoT systems.

The main contributions of this paper are as follows:
\begin{itemize}

\item We conduct an empirical analysis revealing that existing confidence-based EE criteria fail to model the \textit{computational benefit} of deeper inference, which is only realized when EE errors are recoverable by  FE predictions.
\item We propose a novel benefit-aware AMC EE framework, \textit{BEACON}, that shifts the focus from ``prediction confidence" to ``recoverability prediction." At its core, \textit{BEACON} introduces a first-of-its-kind \textit{benefit-aware EE criterion} that explicitly models the potential gain of invoking deeper inference by targeting recoverable errors. 
\item To implement the benefit-aware criterion, we design the LBAP that maps observable probability distribution features from EE branches to a scalar benefit score. This scalar probability score serves as a precise trigger, ensuring that deeper inference is invoked only when a tangible accuracy gain is  expected. The LBAP ensures minimal computational overhead, enabling highly efficient deployment on resource-constrained IoT devices.  
\item Comprehensive evaluations on ResNet-18-based AMC models demonstrate that \textit{BEACON} consistently achieves a superior accuracy-computation trade-off compared with representative baselines. Furthermore, SNR-dependent evaluations confirm that \textit{BEACON} maintains consistent performance gains across varying channel conditions, validating its  robustness and practical suitability  for dynamic wireless environments.

\end{itemize}

The remainder of this paper is organized as follows. Section II  reviews related work on DL-based AMC, EE architectures for resource-efficient AMC, and existing EE criteria. Section III presents the dataset and ResNet-18-based EE model design. Section IV provides an empirical analysis that motivates the proposed framework. In Section V, we introduce the design of LBAP. Experimental results are presented in Section VI, followed by conclusions in Section VII.

\section{Related Work}

\subsection{Deep Learning-Based AMC}
As a fundamental physical-layer task, AMC has long been studied  in wireless communication systems. Early AMC methods primarily relied on likelihood-based or handcrafted feature-based approaches, which often suffer from limited robustness under low SNR conditions and complex channel impairments \cite{zheng2025recent,xu2010lrt_amc}. With the rapid development of data-driven wireless communications \cite{chen2022gp_survey}, DL techniques have been increasingly adopted for AMC to improve classification accuracy and robustness. In particular, CNNs operating directly on raw I/Q samples have demonstrated strong capability in learning discriminative signal representations. Subsequent studies have further enhanced CNN-based AMC models and achieved remarkable performance gains across diverse modulation schemes and SNR regimes \cite{zhang2020cnn, wang2021cnn,liu2022amc, chen2023amc}. 

Despite their superior accuracy and robustness, most existing DL-based AMC models incur high inference complexity. This poses significant challenges for deployment on IoT and edge devices, which typically operate under  limited computational capability and strict energy budget \cite{duan2022distributed,li2025integration,hallaq2025tiny,liu2024user}. This limitation has motivated increasing interest in computation-aware AMC inference mechanisms that aim to reduce inference cost while maintaining classification performance.

\subsection{Early Exit for Resource-Efficient AMC Inference}
Early-exit neural networks have been proposed as an effective approach to reduce inference latency and computational cost by bypassing redundant deeper layers when intermediate predictions meet specific criteria. Representative EE architectures include BranchyNet \cite{teerapittayanon2016branchynet} and shallow-deep networks \cite{kaya2019overthinking}, which introduce auxiliary classifiers at intermediate layers to enable dynamic inference.

Early-exit inference has attracted increasing attention in wireless communication and IoT scenarios, where DL models are often deployed on resource-constrained devices and must satisfy stringent latency and energy requirements. For example, entropy-based EE strategies have been explored for low-precision FPGA implementations, demonstrating notable reductions in computation and energy consumption \cite{Kong2022FPGAEE}. In the AMC context, EE  architectures have been shown to significantly reduce average inference latency without sacrificing overall accuracy, particularly under favorable channel conditions \cite{elsayed2023earlyexit}. Other efforts have further combined EE inference with structured model compression techniques, such as channel pruning, to jointly reduce inference depth and model complexity for AMC in vehicular and edge networks \cite{liu2025joint_earlyexit_pruning}. These studies highlight the promise of EE architectures for resource-efficient AMC deployment.

\subsection{Early-Exit Criteria}

The performance of any EE architecture is fundamentally governed by its exit criterion, which determines whether inference should terminate at an intermediate exit or proceed to deeper network layers. A recent survey \cite{Rahmath2024EarlyExitSurvey} categorizes existing criteria into dynamic (learnable) and static (rule-based) approaches.

Dynamic EE criteria introduce trainable components to adaptively control EE inference decisions based on input features, system constraints, or optimization objectives \cite{chen2020learning,dai2020epnet,bolukbasi2017adaptive}. While flexible, these methods often rely on high-dimensional feature maps or intermediate embeddings, resulting in non-negligible computational and memory overhead.

Static EE criteria, on the other hand, rely on predefined rules derived from EE outputs, typically probability distribution features. Representative static criteria include entropy, maximum softmax probability (MSP), and margin. These methods assess whether the output distribution is sufficiently concentrated, i.e., dominated by a single class, to indicate a confident early prediction. Due to their simplicity and low overhead, entropy-based criteria are predominantly adopted in existing AMC EE studies\cite{elsayed2023earlyexit,liu2025joint_earlyexit_pruning,verbruggen2025deep}.

However, both dynamic and static EE criteria exhibit notable limitations when applied to on-device AMC scenarios. Dynamic methods  introduce additional complexity that may be unsuitable for resource-constrained IoT and edge devices. Static confidence-based criteria, while lightweight, implicitly assume that confidence alone determines the necessity of deeper inference. Consequently, these criteria tend to favor early termination primarily when EE and FE predictions are both correct, without explicitly modeling whether deeper inference would provide actual performance benefit.

To address these limitations, we propose \textit{BEACON}, which explicitly predicts the potential accuracy gain of deeper inference using only low-dimensional EE outputs through a compact predictor. By enabling more informed EE decisions without introducing significant computational or memory overhead, \textit{BEACON} is well suited for deployment on resource-constrained IoT devices.

\begin{figure*}[t]
    \centering
    \includegraphics[width=0.95\linewidth]{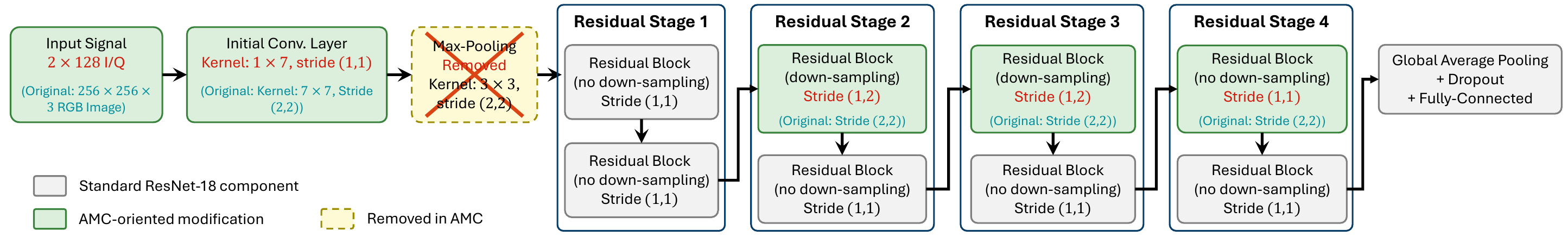}
    \caption{ResNet-18 backbone architecture with AMC-oriented modifications.}
    \label{fig:early_exit_comparison}
    \vspace{-2mm}
\end{figure*}

\section{Early-Exit Model Design for AMC}
In this section, we describe the dataset, backbone architecture, and EE model design adopted for AMC. 
We first introduce the RadioML 2022 (RML22) dataset and the data preprocessing strategy adopted for model training. 
We then present an AMC-oriented modification of the ResNet-18 backbone model to better accommodate complex baseband I/Q signals. 
Finally, we detail the design of the ResNet-18–based AMC EE models, including the EE positions, branch structures, and training strategy.
\subsection{Dataset}
This paper uses the RML22\cite{rml22}, a recently released open-source dataset derived from the  RadioML2016.10A (RML16). RML22 addresses several limitations of RML16 and has been widely adopted as a synthetic benchmark for various wireless communication tasks, including AMC. 
The dataset consists of 420\,000 received baseband signal samples covering ten  modulation schemes: BPSK, QPSK, 8PSK, 16QAM, 64QAM, PAM4, CPFSK, GFSK, WBFM, and AM-DSB. Samples are evenly distributed across 21 SNR levels ranging from $-20$~dB to $20$~dB in steps of $2$~dB, resulting in 2\,000 samples per modulation type per SNR level. Each sample is represented in complex baseband I/Q format with a dimension of $2 \times 128$.
For model development and evaluation, the data is partitioned into training, validation, and test sets using an 81\% / 9\% / 10\% split. This results in 340\,200 training samples, 37\,800 validation samples, and 42\,000 test samples.

To improve generalization, we apply I/Q data augmentation during training. Specifically, we apply stochastic transformations including amplitude scaling, phase rotation, and temporal shifts. Since these transformations preserve the label-defining characteristics of the modulation scheme, they encourage the network to learn distortion-invariant features. This strategy prevents overfitting to specific signal realizations and improves the model's robustness against common wireless impairments.

\subsection{AMC-Oriented ResNet-18 Backbone Model Architecture}

In this paper, we adopt a modified ResNet-18 as the backbone model for AMC. ResNet-based architectures have demonstrated strong classification performance on prior RML benchmarks \cite{abbas2022radio,khan2025automatic}, and ResNet-18 offers a favorable balance between classification accuracy and computational complexity due to its relatively compact structure.

However, the standard ResNet-18 is optimized for image data, whereas complex baseband I/Q signals possess distinct temporal and channel-wise characteristics. As illustrated in Fig.~\ref{fig:early_exit_comparison}, we introduce several AMC-oriented structural modifications on the original ResNet-18 architecture:
\begin{itemize}
    \item \textbf{Input Layer Adaptation:} The initial $7\times7$  convolution with stride $(2,2)$ is replaced by a $1\times7$ convolution with stride $(1,1)$ to enable temporal-only feature extraction without early down-sampling.
    \item \textbf{Information Preservation:} The subsequent max-pooling layer is removed to prevent premature information loss, which is critical given the relatively short $128$-sample temporal window.
    \item \textbf{Strategic Down-sampling:} We restrict down-sampling exclusively to the temporal dimension across the residual stages. In Residual Stage 1, we maintain full resolution using a $(1,1)$ stride for both basic blocks. In Residual Stages~2 and 3, we apply a $(1,2)$ stride in their first blocks to reduce the temporal dimensionality from 128 to 64 and from 64 to 32, respectively. In Residual Stage~4, no additional down-sampling is applied. Both blocks operate with stride $(1,1)$ to maintain sufficient temporal resolution for deeper feature representations.
    \item \textbf{Classification Head}: Global features are aggregated via adaptive average pooling followed by a fully connected layer for modulation classification.
\end{itemize}

The modified ResNet-18 achieves approximately 71\% classification accuracy on both the validation and test sets, outperforming the original ResNet-18 baseline (63\%) and demonstrating competitive performance compared with recently reported models on the RML22 dataset \cite{wang2025complexhybrid}.

\begin{figure}[t]
    \centering
    \includegraphics[width=0.8\linewidth]{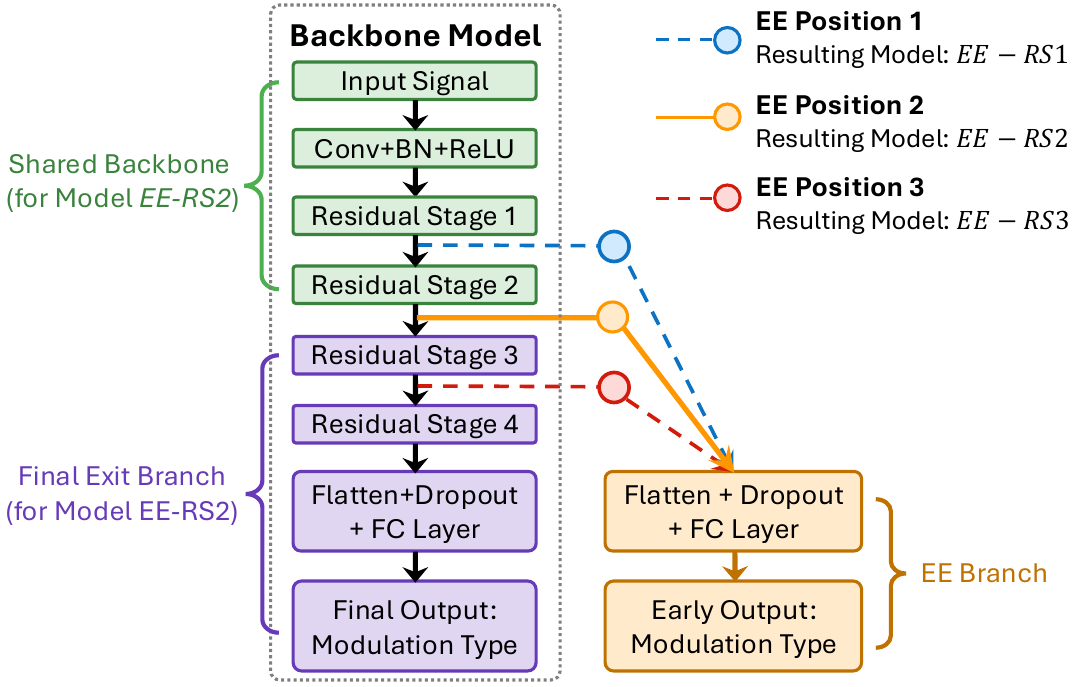}
    \caption{Early-exit configurations in the  AMC-oriented ResNet-18 backbone.}
 \label{fig:early_exit_arch}
\end{figure}

\subsection{ResNet-18-Based EE Model Design}
To explore the trade-off between computational latency and classification accuracy, we implement three EE configurations based on the modified ResNet-18 backbone. The resulting models, denoted as \textit{EE-RS1}, \textit{EE-RS2}, and \textit{EE-RS3}, correspond to attaching the EE branch after Residual Stages 1, 2, and 3, respectively. As illustrated in Fig.~\ref{fig:early_exit_arch}, 
all three variants share the same backbone architecture and identical EE branch design, differing solely in their EE branch attachment positions.

Following established EE training practices \cite{xin2021berxit, gong2023trainingexit}, we employ a decoupled training strategy. The backbone parameters are first optimized to convergence and subsequently frozen. 
For each EE configuration, the EE branch is trained independently to minimize interference with the FE optimization and  ensure stable convergence of auxiliary classifiers. 

We define two critical reference metrics to evaluate the EE framework: 1) \textit{Backbone Accuracy:} The classification accuracy obtained when all test samples are forwarded to the FE branch, representing the upper bound of accuracy and computation cost; 2) \textit{EE Branch Accuracy:} The accuracy achieved when all test samples terminate at the EE branch, representing the lower bound of accuracy and minimum computation cost. 
These two reference accuracies provide quantitative baseline for evaluating the trade-off between computational cost and classification performance under different EE strategies.

\section{Analysis and Motivation for Benefit-Aware EE}\label{sec_motivation}

Most existing EE methods for wireless communication applications rely on confidence-based criteria, such as entropy, MSP, or margin. These methods are attractive due to their effectiveness, simplicity, and low computational overhead. 
In this work, we conduct rigorous empirical analysis on ResNet-18-based AMC EE models to evaluate whether entropy, the most widely adopted EE criterion,  truly reflects the benefit of deeper inference.
To facilitate the analysis, we first introduce an entropy-bin-based evaluation framework and an EE outcome taxonomy. 
Using this framework, we derive two key observations that reveal fundamental limitations of entropy-based early exiting, motivating the need for a benefit-aware EE criterion.

\begin{table*}[t]
\centering
\caption{Per-Entropy-Bin Distribution of EE and FE Outcomes on Model \textit{EE-RS2}.}
\setlength{\tabcolsep}{12pt}
\label{tab:entropy_bin_4case_v2}
\begin{tabular}{lccccc}
\toprule
Entropy bin &
Samples (\%) &
\textbf{Case $\mathcal{C}_{11}$} (\%) &
\textbf{Case $\mathcal{C}_{01}$}  (\%) &
\textbf{Case $\mathcal{C}_{00}$} (\%) &
\textbf{Case $\mathcal{C}_{10}$}  (\%) \\
\midrule
$[0.0,0.1)$ & 0.00 & 100.00 & 0.00 & 0.00 & 0.00 \\
$[0.1,0.2)$ & 0.12 & 98.08 & 0.00 & 1.92 & 0.00 \\
$[0.2,0.3)$ & 2.99 & 87.16 & 3.51 & 9.33 & 0.00 \\
$[0.3,0.4)$ & 5.24 & 81.23 & 11.64 & 6.91 & 0.23 \\
$[0.4,0.5)$ & 6.13 & 92.35 & 5.24 & 2.29 & 0.12 \\
$[0.5,0.6)$ & 9.61 & 97.25 & 1.78 & 0.94 & 0.02 \\
$[0.6,0.7)$ & 5.07 & 96.25 & 1.97 & 1.69 & 0.09 \\
$[0.7,0.8)$ & 16.00 & 52.30 & 46.21 & 1.06 & 0.43 \\
$[0.8,0.9)$ & 13.44 & 53.61 & 35.30 & 5.83 & 5.26 \\
$[0.9,1.0]$ & 41.40 & 16.42 & 19.18 & 55.29 & 9.11 \\
\bottomrule
\end{tabular}
\vspace{-2mm}
\end{table*}

\subsection{Entropy Bins and Early-Exit Taxonomy}

To investigate whether EE entropy reliably reflects the potential benefit of  deeper inference, we conduct a \textbf{per-entropy-bin analysis} on the test set. Specifically, for each test sample, we first compute the normalized entropy of the EE probability distribution. Given the predicted class probability vector $\mathbf{p}_e=[p_{e,1},\dots,p_{e,C}]$ produced by the EE classifier, the entropy score is defined as:
\begin{equation}\label{eq1}
\mathcal{S}_{\mathrm{ent}}(\mathbf{p}_e)
= \frac{- \sum_{c=1}^{C} p_{e,c} \log p_{e,c}}{\log C},
\end{equation}
where $C$ denotes the number of classes and $p_{e,c}$ is the probability of class $c$ produced by the EE branch. The score is normalized to the range $[0, 1]$. Low entropy indicates that the probability mass is concentrated on a single class (high prediction confidence), while high entropy corresponds to a more uniform distribution and higher prediction uncertainty.

Samples are then partitioned into discrete bins of width 0.1 based on their entropy values. 
For analytical purposes, all samples are forwarded through both the EE branch and the FE branch, regardless of any early-stopping decision. This allows us to compare EE and FE predictions for the same input.

Based on the EE and FE predictions, we categorize each sample into one of four mutually exclusive cases:
\begin{itemize}
\item \textbf{Case $\mathcal{C}_{11}$ (Consistent Corrects):} Predictions from both exits are correct; deeper inference is redundant.
    \item \textbf{Case $\mathcal{C}_{01}$ (Recoverable Errors):} The EE prediction is wrong, but the final exit corrects it. \textit{This is the only case where deeper inference provides a functional benefit.}
    \item \textbf{Case $\mathcal{C}_{00}$ (Irrecoverable Errors):} Both exits are wrong. Deeper inference wastes energy without gaining accuracy.
    \item \textbf{Case $\mathcal{C}_{10}$ (Prediction Degradation):} The EE prediction is correct, but deeper inference is wrong, i.e., a phenomenon often termed ``overthinking" \cite{kaya2019overthinking}.
\end{itemize}

Table~\ref{tab:entropy_bin_4case_v2} presents the distribution of these cases across entropy bins for the AMC model \textit{EE-RS2}. The ``Samples (\%)'' column indicates the fraction of test samples in each entropy bin relative to the entire test set. 
All other percentages are calculated relative to the sample count within each bin.

\subsection{Observation 1: Entropy Does Not Reliably Indicate the Benefit of Deeper Inference}

Under conventional entropy-based EE, samples with entropy below a predefined threshold are exited early.
As shown in Table~\ref{tab:entropy_bin_4case_v2}, low-entropy bins  are dominated by Case $\mathcal{C}_{11}$ (Consistent Corrects).
This  aligns with the underlying assumption of all confidence-based criteria: a concentrated output distribution dominated by a single class indicates high model confidence, making an early exit safe.

However, these confident samples (entropy below 0.7) constitute less than 30\% of the entire test set. The remaining majority falls into higher-entropy bins (e.g., $[0.7, 1.0]$). In these regions, the relationship between entropy and the necessity of further computation becomes decoupled:
\begin{itemize}
    \item \textbf{Redundant Computation:} Many high-entropy samples fall into Case $\mathcal{C}_{00}$, $\mathcal{C}_{11}$ and $\mathcal{C}_{10}$, where deeper inference is either ineffective ($\mathcal{C}_{00}$ and $\mathcal{C}_{11}$) or even harmful ($\mathcal{C}_{10}$). For these samples, early termination is preferable.
    \item \textbf{Missed Recoverable Gains:} While some studies \cite{verbruggen2025widthwise,verbruggen2025deep} suggest terminating high-entropy samples to avoid wasted effort on irrecoverable errors ($\mathcal{C}_{00}$), our results show that approximately $46\%$, $35\%$, and $19\%$ of samples in the entropy bins $[0.7,0.8)$, $[0.8,0.9)$, and $[0.9,1.0]$, respectively, are recoverable errors ($\mathcal{C}_{01}$). Since these recoverable errors constitute about 21\% of the total test set and are primarily located in high-entropy bins, aggressively terminating uncertain samples would forfeit non-negligible accuracy gains achievable through deeper inference. 
\end{itemize}

These results expose a fundamental limitation of entropy-based EE criteria. While entropy effectively measures prediction confidence, it does not directly indicate whether deeper inference is beneficial. Therefore, entropy alone is insufficient for principled, benefit-aware EE decisions in AMC.

\subsection{Observation 2: Class-Specific Structural Information Loss in Confidence-Based Criteria}

\begin{figure}[t]
    \centering
    
    \begin{subfigure}{0.45\textwidth}
        \centering
        \includegraphics[width=0.8\linewidth]{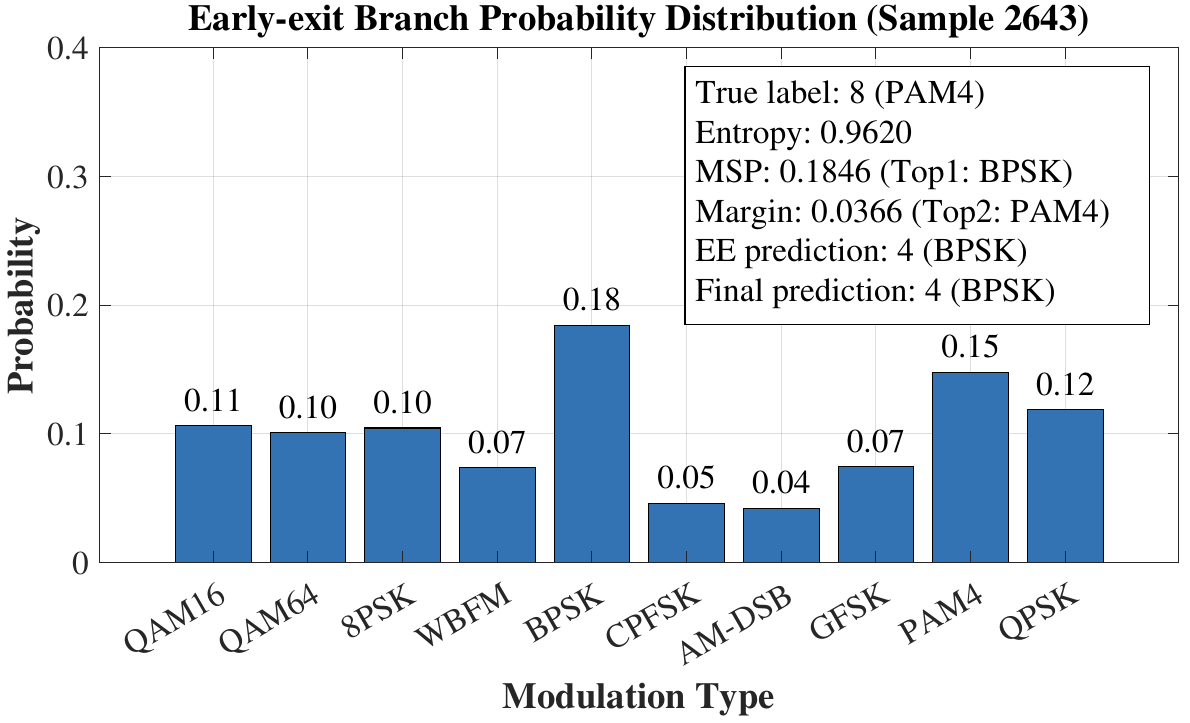}
        
        \vspace{-2mm}
        \captionsetup{font=scriptsize}
        \caption{\textbf{Case $\mathcal{C}_{00}$ (Irrecoverable Error)}: Competing Classes BPSK and PAM4}
        \label{fig4-1}
    \end{subfigure}
    
    \hfill
    
    \vspace{-1mm}
    \begin{subfigure}{0.45\textwidth}
        \centering
        \includegraphics[width=0.8\linewidth]{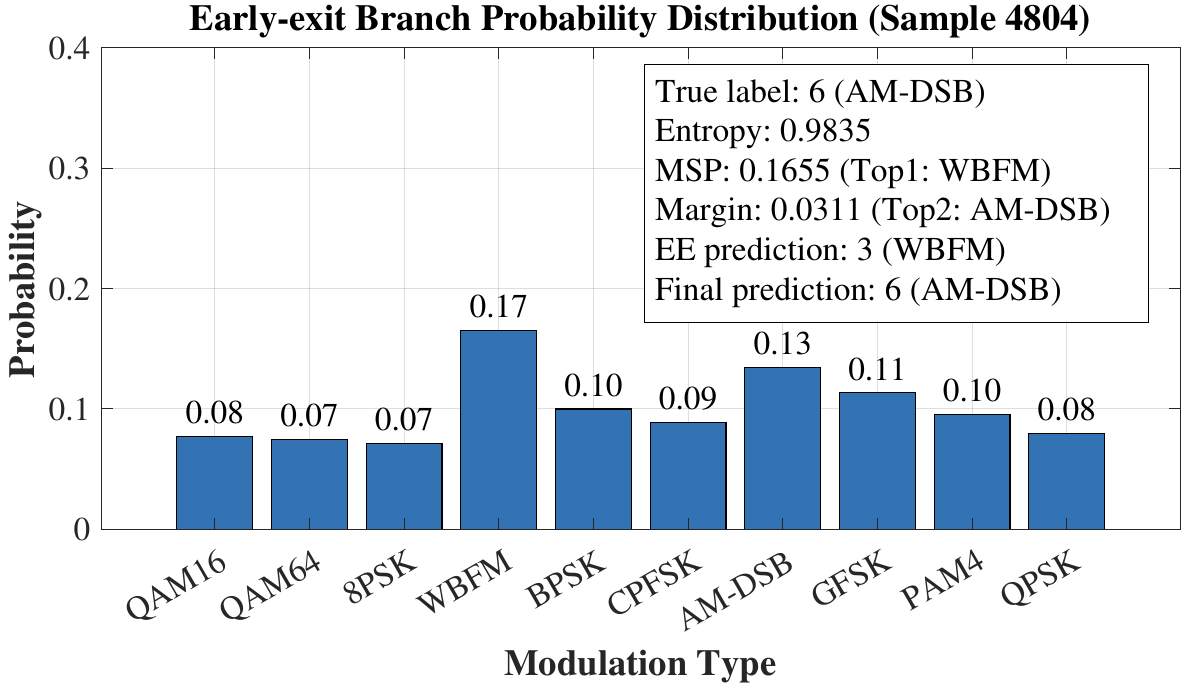}

        \vspace{-2mm}
        \captionsetup{font=scriptsize}
        \caption{\textbf{Case $\mathcal{C}_{01}$ (Recoverable Error)}: Competing Classes WBFM and AM-DSB}
        \label{fig4-2}
    \end{subfigure}
    
    \caption{Two EE output probability distributions with similar confidence statistics but different recoverability outcomes.}
    \label{fig4}
\end{figure}

Through our empirical analysis, we observe that samples  with nearly identical confidence statistics can exhibit completely different recoverability behavior. Fig.~\ref{fig4} shows two representative samples with highly similar entropy, MSP, and margin values. Their output distributions are similarly flat, each with two classes slightly dominating the others. Under conventional scalar confidence criteria, these samples are mathematically indistinguishable. A threshold-based mechanism would therefore treat them identically, either exiting or continuing both, despite their different potential for recovery.

Despite their statistical similarity, their final outcomes are fundamentally different. In Fig.~\ref{fig4-1}, both  EE and FE predictions are incorrect, representing an irrecoverable error (Case $\mathcal{C}_{00}$). In contrast, Fig.~\ref{fig4-2} shows a recoverable error (Case $\mathcal{C}_{01}$) where the incorrect EE prediction is  corrected at the FE branch. 

A deeper analysis of the full probability distributions reveals the key distinction: \textit{class identity matters}. In Fig.~\ref{fig4-1}, the ambiguity is between BPSK and PAM4, a confusion that persists through deeper layers. Conversely, the competition in Fig.~\ref{fig4-2} involves WBFM and AM-DSB, a pair that the model can more effectively disentangle with further inference.  

More generally, the recoverability of an EE prediction error depends not only on the degree of uncertainty, but also on the specific classes competing in the distribution. Certain inter-class confusions are inherently easier to resolve through deeper inference than others. 
Confidence-based criteria, however, compress the entire probability distribution into  scalar measures (e.g., entropy, MSP, or margin). This process discards \textit{class-specific structural information}, treating distinct confusion patterns as equivalent whenever overall confidence levels are similar. Consequently, such criteria cannot differentiate between a ``difficult but recoverable" signal and an ``irrecoverable" one. This observation motivates leveraging the full EE probability distribution to capture class-specific patterns and make more informed, benefit-aware EE decisions.

\section{Proposed \textit{BEACON} Framework}

To address the limitations of confidence-based criteria discussed in Section \ref{sec_motivation}, we propose \textit{BEACON}, a unified benefit-aware early-exit framework. Instead of relying on confidence measures, \textit{BEACON} directly models the expected accuracy gain from deeper inference. \textit{BEACON} consists of two core components: a formal benefit-aware recoverability criterion and a lightweight predictor p designed for real-time execution.

\subsection{Benefit-Aware Recoverability Criterion}
Consider an AMC backbone model with an EE branch and a FE branch (e.g., models \textit{EE-RS1}, \textit{EE-RS2}, or \textit{EE-RS3}). With an input signal sample $\mathbf{x}$, let $\mathbf{z}_e(\mathbf{x}), \mathbf{z}_f(\mathbf{x}) \in \mathbb{R}^C$ denote the logits produced by the EE and FE branches, respectively, 
where $C$ is the number of modulation classes.
The corresponding softmax probability vectors are defined as $\mathbf{p}_e = \sigma(\mathbf{z}_e)$ and $\mathbf{p}_f = \sigma(\mathbf{z}_f)$, yielding the predicted labels:
\begin{equation}
\hat{y}_e = \arg\max_c p_{e,c}, \quad
\hat{y}_f = \arg\max_c p_{f,c},
\end{equation}
where $p_{e,c}$ and $p_{f,c}$ denote the probabilities of class $c$ produced by the EE branch and the FE branch, respectively.

The core innovation of \textit{BEACON} lies in identifying  \textit{recoverable errors} ($\mathcal{C}_{01}$).
Given the ground-truth label $y$, we define the recoverability indicator $\mathbb{I}_{\mathrm{rec}}(\mathbf{x})$ as
\begin{align}\label{eq3}
\mathbb{I}_{\mathrm{rec}}(\mathbf{x}) =
\begin{cases}
1, & \text{if } \hat{y}_e \neq y \text{ and } \hat{y}_f = y \\
0, & \text{otherwise}
\end{cases}
\end{align}
\textit{BEACON} defines a benefit-aware criterion $\mathcal{R}$ to estimate the conditional probability that a sample belongs to Case  $\mathcal{C}_{01}$:
\begin{align}
    \mathcal{S}_\mathcal{R}(\mathbf{p}_e)=P(\mathbb{I}_{\mathrm{rec}}=1 \mid \mathbf{p}_e). 
\end{align}
Unlike existing methods that exit based on local certainty, $\mathcal{S}_\mathcal{R}$ directly estimates whether forwarding a sample to deeper layers is beneficial. This criterion ensures that computational resources are preserved for instances where deeper inference provides a tangible accuracy gain.

\subsection{Lightweight Benefit-Aware Predictor Design}
\begin{figure*}[t]
    \centering
    \includegraphics[width=0.9\linewidth]{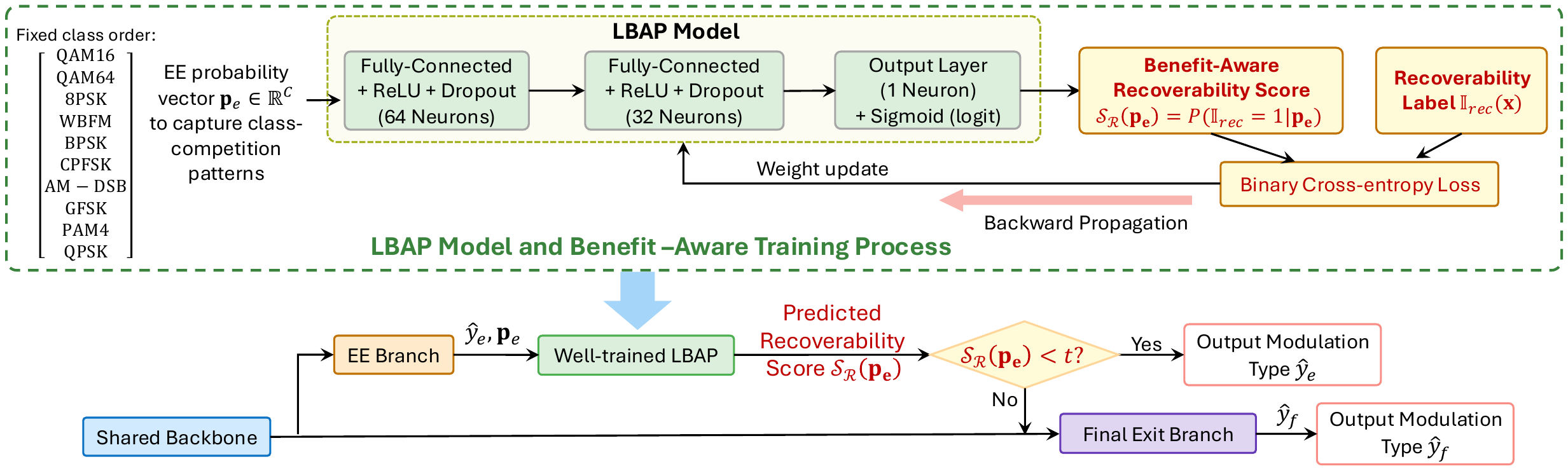}
    \caption{Proposed \textit{BEACON}-based EE framework for AMC.}
    \label{fig:LBAP}
\end{figure*}

\begin{table}[t]
\centering
\caption{Computational and Parameter Overhead of LBAP.}
\label{tab:overhead}
\begin{tabular}{lcc}
\toprule
\textbf{Module} & \textbf{MACs} & \textbf{Parameters} \\
\midrule
Backbone (ResNet-18) 
& 388,929,024 
& 11,173,130 \\

LBAP 
& 2,720 (0.0007\%) 
& 2,817 (0.0252\%) \\
\bottomrule
\end{tabular}
\end{table}
To realize the recoverability criterion in a real-time AMC system, we design the LBAP to estimate $\mathcal{S}_\mathcal{R}$ using only the EE probability distribution, as shown in Fig.~\ref{fig:LBAP}.

\subsubsection{Input Representation} Instead of compressing $\mathbf{p}_e$ into a scalar statistic,  LBAP processes the full probability vector
\begin{align}
    \mathbf{p}_e = [p_{e,1}, p_{e,2}, \cdots, p_{e,C}]^{\top}.
\end{align}
In our AMC setting, $C = 10$ and $c=1, 2, \cdots 10$ correspond to the following fixed modulation order:
\begin{align}
    \{&1: \text{QAM16}, 2: \text{QAM64}, 3: \text{8PSK}, 4: \text{WBFM}, 5: \text{BPSK}, \nonumber \\
&6: \text{CPFSK}, 7: \text{AM-DSB}, 8: \text{GFSK}, 9: \text{PAM4}, 10: \text{QPSK}\}\nonumber 
\end{align}
The class ordering is fixed throughout training and inference.
By processing the entire structured probability vector, the model can capture structural information and learn  class-dependent competition structures embedded in the distribution.

\subsubsection{Model Architecture} 
To ensure deployment feasibility on resource-constrained  edge nodes,  LBAP is implemented as a compact multi-layer perceptron with two fully connected hidden layers with 64 and 32 neurons, respectively. 
Each hidden layer is followed by a ReLU activation and dropout for regularization. 
The final layer outputs a single scalar logit, $f(\mathbf{p}_e)$, through a single neuron with a Sigmoid activation $s(\cdot)$, mapping the latent features to the benefit score (estimated recoverable probability) $\mathcal{S}_\mathcal{R}(\mathbf{p}_e)=s\!\left(f(\mathbf{p}_e)\right) \in [0,1]$.

\subsubsection{Training Strategy}
The LBAP is trained to approximate the benefit-aware criterion through supervised learning. Each sample $\mathbf{x}$ is assigned a binary recoverability label $\mathbb{I}_{\mathrm{rec}}(\mathbf{x})$.  
LBAP is trained using the binary cross-entropy loss:
\begin{equation}
\mathcal{L} = - \mathbb{I}_{\text{rec}} \log \mathcal{S}_\mathcal{R}(\mathbf{p}_e)
- (1 - \mathbb{I}_{\text{rec}}) \log \left(1 - \mathcal{S}_\mathcal{R}(\mathbf{p}_e)\right).
\end{equation}
Note that LBAP is trained post-hoc while freezing  backbone and EE branch parameters. 
This preserves the original classification behavior while layering \textit{BEACON} as a lightweight decision module on top of a pretrained AMC network.

\subsubsection{Complexity Analysis}
As shown in Table~\ref{tab:overhead}, LBAP introduces 
only 2,720 multiply–accumulate operations (MACs) ($\approx$ 0.0007\% of the ResNet-18 backbone) and 2,817 trainable parameters ($\approx$  0.025\% of the backbone size).
This negligible overhead enables seamless integration into practical AMC systems without affecting real-time deployment constraints.

\subsection{\textit{BEACON}-Based Inference and Decision Logic}

\subsubsection{Overall Decision Mechanism}
The overall \textit{BEACON}-based inference process is illustrated in Fig.~\ref{fig:LBAP}. 
Given an input sample: 1) The shared backbone  extracts intermediate features; 2) The EE branch produces a prediction along with its corresponding probability vector $\mathbf{p}_e$; 3) The probability vector is  fed into LBAP, which outputs a  recoverability score $\mathcal{S}_\mathcal{R}(\mathbf{p}_e)$. 

We employ a threshold-based decision rule to govern the inference flow. 
If $S_\mathcal{R}(\mathbf{p}_e)$ is lower than a predefined threshold $t \in (0,1)$, the sample is deemed unlikely to benefit from deeper inference, and the EE prediction is thus accepted as the final output. 
Otherwise, samples with $S_\mathcal{R}(\mathbf{p}_e)\geq t$ are forwarded to the FE branch for full-depth inference. This preserves the classic \emph{score-and-threshold} paradigm  while replacing uncertainty scores with a benefit-aware score. \textit{BEACON}-based EE process is summarized in Algorithm~\ref{alg:benefit_exit}. 

\subsubsection{Accuracy-Efficiency Tradeoff Analysis}
The threshold $t$ directly controls the balance between computational efficiency and prediction accuracy. 
A high threshold ($t\to1$) prioritizes computational efficiency. Only samples with a high recoverability probability proceed to the FE branch. While this minimizes redundant computation, it may lead to false negatives where potentially correctable errors are terminated early. 
Conversely, a low threshold ($t\to0$) prioritizes inference accuracy. By allowing more samples to reach the FE branch, the system maximizes the opportunity for error correction at the expense of increased latency and power consumption.
Consequently, by adaptively adjusting the threshold $t$, \textit{BEACON} enables a tunable mechanism to balance accuracy and computational cost according to diverse deployment requirements.

\begin{algorithm}[!t]
\caption{\textit{BEACON}-Based AMC EE Mechanism}
\label{alg:benefit_exit}
\KwIn{Dataset $\mathcal{D}\! =\! \{\! (x_i,y_i)\! \}_{i=1}^{N}$; shared backbone~$\phi(\! \cdot\! )$; 
EE branch $b_e(\cdot)$; FE branch $b_f(\cdot)$; LBAP predictor $s(f_\theta(\cdot))$; threshold $t$; 
learning rate $\eta$; training epochs $E$}
\KwOut{Trained predictor parameters $\theta$; inference rule for predicted label $\hat{y}$}

\textbf{\textit{Offline Training Stage:} Train LBAP $s\!\left(f_\theta(\cdot)\right)$ with frozen backbone/exits}\;
 Freeze parameters in shared backbone $\phi(\cdot)$, EE branch $b_e(\cdot)$, and FE branch $b_f(\cdot)$. \\
 Initialize LBAP parameters $\theta$. \\
\For{$epoch=1$ \KwTo $E$}{
  \For{each mini-batch $\mathcal{B}\subset\mathcal{D}$}{
  Initialize accumulated batch loss: $\mathcal{L} \leftarrow 0$
  
    \For{each $(x,y)\in\mathcal{B}$}{
      Extract features via shared backbone: $\mathbf{f} \leftarrow \phi(\mathbf{x})$. \\
            Produce EE and FE logits: $\mathbf{z}_e \leftarrow b_e(\mathbf{f}), \;\; \mathbf{z}_f \leftarrow b_f(\mathbf{f})$. \\
            Compute EE and FE probability vectors: $\mathbf{p}_e \leftarrow \sigma(\mathbf{z}_e), \;\; \mathbf{p}_f \leftarrow \sigma(\mathbf{z}_f)$. \\
            Obtain EE and FE predicted labels: $\hat{y}_e \leftarrow \arg\max_c p_{e,c},\;\; \hat{y}_f \leftarrow \arg\max_c p_{f,c}$. \\
            Generate benefit-aware recoverability indicator $\mathbb{I}_{\mathrm{rec}}$, based on Eq. \eqref{eq3}, as training labels. \\
            Predict recoverability benefit score from EE output: $\mathcal{S}_\mathcal{R}(\mathbf{p}_e) \leftarrow s\!\left(f_\theta(\mathbf{p}_e)\right)$. \\
            Calculate binary cross-entropy loss: $\ell \leftarrow -\mathbb{I}_{\mathrm{rec}}\log(\mathcal{S}_\mathcal{R}(\mathbf{p}_e))-(1-\mathbb{I}_{\mathrm{rec}})\log(1-\mathcal{S}_\mathcal{R}(\mathbf{p}_e))$. \\
            Accumulate loss: $\mathcal{L} \leftarrow \mathcal{L}+\ell$
    }
    Update LBAP parameters: $\theta \leftarrow \theta - \eta \nabla_\theta
    \frac{1}{|\mathcal{B}|}\mathcal{L}$\;
  }
}

\textbf{\textit{Online Inference Stage:} Benefit-aware inference with trained LBAP $s(f_\theta(\cdot))$}\;

For input sample $\mathbf{x}$, compute recoverability score based on EE probability vector $\mathbf{p}_e$:
$\mathcal{S}_\mathcal{R}(\mathbf{p}_e) \leftarrow s\!\left(f_\theta(\mathbf{p}_e)\right)$. \\
\eIf{$\mathcal{S}_\mathcal{R}(\mathbf{p}_e) < t$}{
    \Return EE prediction $\hat{y} \leftarrow \arg\max_c p_{e,c}$ 
}{
    Perform full inference to get FE probability vector $\mathbf{p}_f$.\\
    \Return FE prediction $\hat{y} \leftarrow \arg\max_c p_{f,c}$ 
}
\vspace{-2mm}
\end{algorithm}

\section{Experimental Results and Analysis}

\subsection{Experimental Setup}
All experiments are conducted on a workstation equipped with an NVIDIA GeForce RTX~3050 laptop GPU with 6~GB VRAM, using the PyTorch framework. We evaluate five  representative confidence-based and two learning-based EE criteria as baselines. Confidence-based EE criteria include entropy (the predominant choice in wireless communication scenarios), MSP, margin, top-3 mass sum, and the Gini index, which are widely used in EE and selective inference studies across other application domains \cite{Rahmath2024EarlyExitSurvey}. Two learning-based criteria, namely the learning-based confidence predictor (LCFP) and the learning-based consistency predictor (LCSP), use the same LBAP architecture to ensure a fair comparison. LCFP estimates prediction confidence, with the same objective as conventional static confidence-based criteria, but implemented in a learning-based manner. In contrast, LCSP predicts whether the EE prediction matches the FE prediction, an approach widely adopted in NLP-based EE frameworks \cite{xin2021berxit,zhou2020bert,zhu2021gamlbert}.

To enable fair comparison, we adopt a unified score-and-threshold framework. 
Let $\mathbf{p}_e = [p_{e,1}, \dots, p_{e,C}]^{\top}$ denote the softmax probability vector at the EE branch. 
Each criterion produces a scalar score $\mathcal{S}(\mathbf{p}_e)$. To ensure consistency with the \textit{BEACON} benefit score $\mathcal{S}_\mathcal{R}(\mathbf{p}_e)$, we define all scoring functions such that larger scores consistently indicate higher uncertainty, and thus a higher likelihood that the sample requires further processing by the FE branch. For criteria that inherently assign larger values to more confident predictions (MSP, Margin, Top-$k$, LCFP, and LCSP), we apply a monotonic transformation (i.e., $1-\cdot$) to align them with this uncertainty-based definition. Under this framework, a sample is early-exited if $\mathcal{S}(\mathbf{p}_e) < t$; otherwise, it is forwarded to the FE branch. The specific scoring functions and their corresponding score ranges are defined as follows: 
\begin{itemize}
\item \textbf{Entropy:} $\mathcal{S}_{\mathrm{ent}}(\mathbf{p}_e) \in [0,1]$ as defined in Eq. \eqref{eq1}. 
\item \textbf{MSP:} $\mathcal{S}_{\mathrm{msp}}(\mathbf{p}_e)
= 1 - \max_{c} \; p_{e,c}$, with score range $\mathcal{S}_{\mathrm{msp}}(\mathbf{p}_e) \! \in \! \left[0, 1-\frac{1}{C}\right]$
\item \textbf{Margin:} $\mathcal{S}_{\mathrm{mrg}}(\mathbf{p}_e)
= 1 - (p_{e,(1)} - p_{e,(2)})$, where $p_{e,(1)}$ and $p_{e,(2)}$ denote the largest and second-largest elements of $\mathbf{p}_e$, respectively. The score range is $\mathcal{S}_{\mathrm{mrg}}(\mathbf{p}_e) \in [0,1]$.
\item \textbf{Top-$k$ Mass Sum} (Top-3, $k\! =\! 3$ in our experiments): $\mathcal{S}_{\mathrm{top}\text{-}k}(\mathbf{p}_e)
= 1 - \sum_{i=1}^{k} p_{e,(i)}$, where $p_{e,(i)}$ denotes the $i$-th largest probability in $\mathbf{p}_e$. The score range is $\mathcal{S}_{\mathrm{top}\text{-}3}(\mathbf{p}_e)\! \in \! \left[0, 1-\frac{3}{C}\right]$
\item \textbf{Gini Index:} $\mathcal{S}_{\mathrm{gini}}(\mathbf{p}_e)
= 1 - \sum_{c=1}^{C} p_{e,c}^{2}$, with score range $\mathcal{S}_{\mathrm{gini}}(\mathbf{p}_e) \in \left[0, 1-\frac{1}{C}\right]$

\item \textbf{LCFP:} $\mathcal{S}_{\mathrm{lcfp}}(\mathbf{p}_e) \in [0,1]$ is the predicted confidence level, estimating the probability of a correct EE output.
\item \textbf{LCSP:} $\mathcal{S}_{\mathrm{lcsp}}(\mathbf{p}_e) \in [0,1]$ estimates the probability that the EE prediction is inconsistent with the FE prediction.
\end{itemize}

In practice, EE thresholds are typically manually tuned  within the valid  range of each score. 
To avoid bias introduced by static, manually-selected threshold and to rigorously characterize the accuracy-computation trade-off, we adopt a percentile-based threshold sweeping strategy. Specifically, for each criterion, we evaluate 21 operating points corresponding to percentiles from 0\% to 100\% with a step size of 5\%.
The two endpoints define the extremes: 100\% corresponds to always accepting the EE prediction, whereas 0\% disables early exit entirely (i.e., forwarding all samples to the FE branch). 
By incrementally relaxing the threshold, we track performance as more samples  terminate inference at the EE branch.

\begin{table}[t]
\centering
\caption{Performance Statistics for AMC EE Models}
\label{tab:ee_basic_stats}
\resizebox{\linewidth}{!}{%
\begin{tabular}{lcccc}
\toprule
\textbf{Model} 
& \textbf{EE Acc. (\%)} 
& \textbf{FE Acc. (\%)} 
& \textbf{$P_{recov}$ (\%)} 
& \textbf{$P(\mathcal{C}_{01} \! \mid \! \mathcal{C}_{01} \cup \mathcal{C}_{00})$ (\%)} \\
\midrule
\textit{EE-RS1} & 26.29 & 70.32 & 46.71 & 63.38 \\
\textit{EE-RS2} & 53.73 & 70.32 & 21.24 & 45.90 \\
\textit{EE-RS3} & 67.36 & 70.32 & 6.62 & 20.29 \\
\bottomrule
\end{tabular}%
}
\end{table}

\subsection{Accuracy and Recoverability Statistics of AMC EE Models}

We evaluate the seven baseline criteria across three AMC EE models (i.e., \textit{EE-RS1}, \textit{EE-RS2}, and \textit{EE-RS3} with different EE positions). For fairness, each criterion is evaluated independently under each EE configuration.
 Table~\ref{tab:ee_basic_stats} summarizes the EE and FE accuracy statistics on the validation dataset. As shown in the table, while FE accuracy remains invariant across all models due to shared backbone parameters, EE accuracy increases with depth, due to stronger feature representations with deeper layers. 
 
The recoverable rate $P_{recov}$, defined as the proportion of Case $\mathcal{C}_{01}$ samples within the total validation set, is a key quantity in our analysis. As the EE branch moves deeper, $P_{recov}$ decreases significantly due to two factors. First, deeper EE branches are more accurate, leaving fewer EE misclassifications that need correction. Second, as the EE branch moves closer to the FE, the reduced depth gap leaves less representational capacity for correcting errors. This effect is further reflected by the conditional probability $P(\mathcal{C}_{01} \! \mid \! \mathcal{C}_{01} \cup \mathcal{C}_{00})$, which measures the probability that an early misclassification can be corrected by the FE branch. This probability falls from 63.38\% (\textit{EE-RS1}) to 20.29\% (\textit{EE-RS3}), indicating the diminishing corrective capacity of the remaining network  as the depth gap narrows.

\subsection{Computational Cost-Accuracy Trade-off}
\begin{figure*}[t]
    \centering
    \subfloat[\footnotesize Model \textit{EE-RS1}]{
        \includegraphics[width=0.31\linewidth]{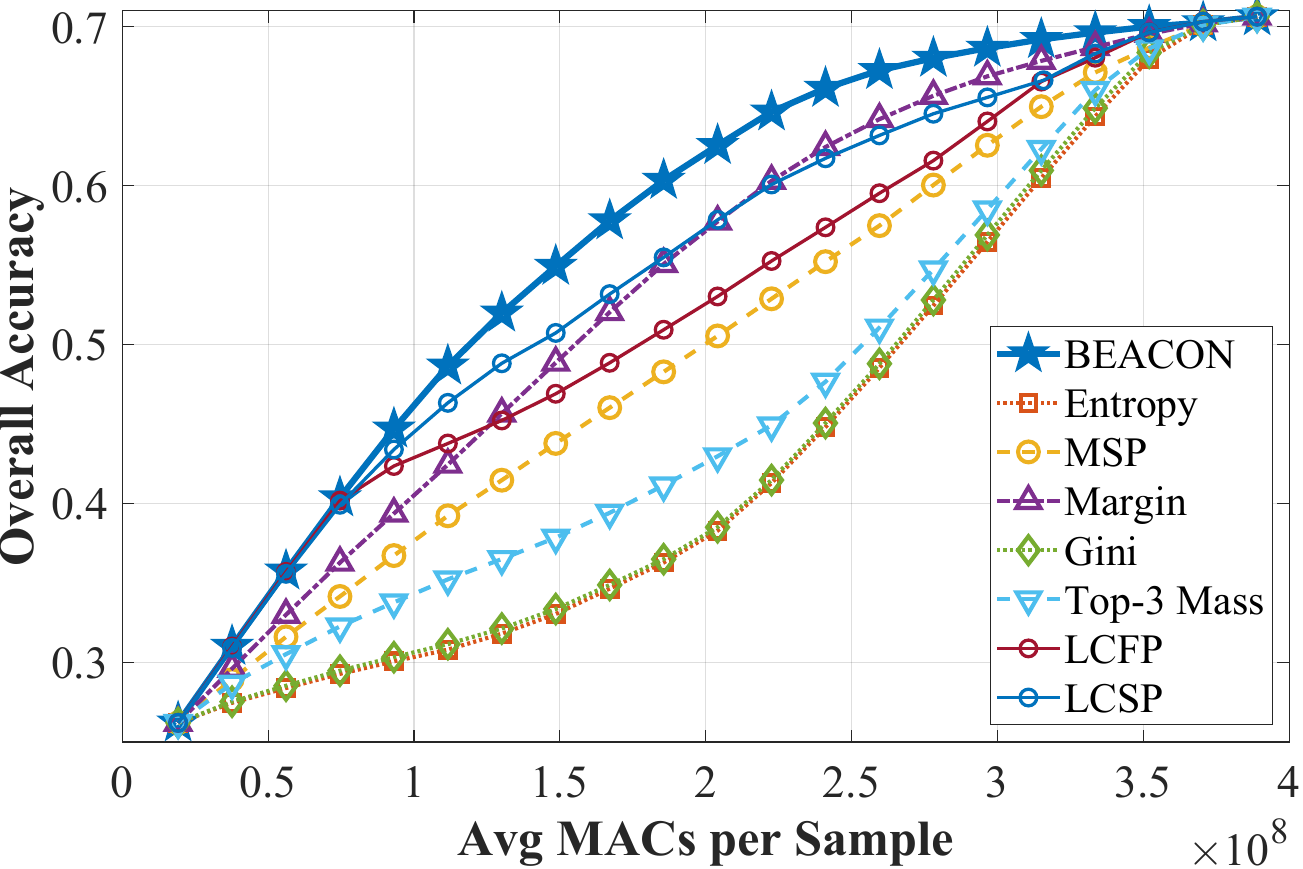}
        \label{fig:tradeoff_eel1}
    }
    \hfill
    \subfloat[\footnotesize Model \textit{EE-RS2}]{
        \includegraphics[width=0.31\linewidth]{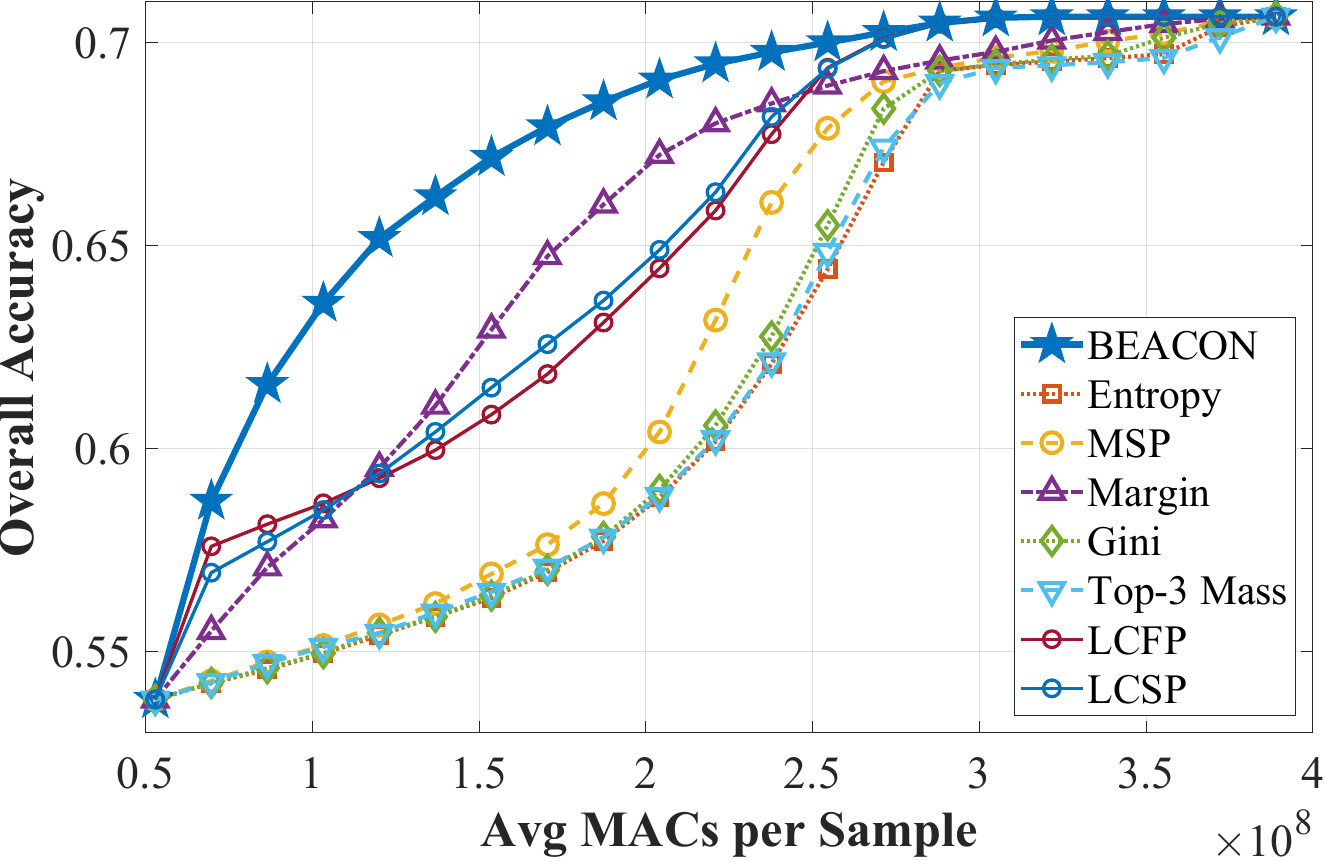}
        \label{fig:tradeoff_eel2}
    }
    \hfill
    \subfloat[\footnotesize Model \textit{EE-RS3}]{
        \includegraphics[width=0.31\linewidth]{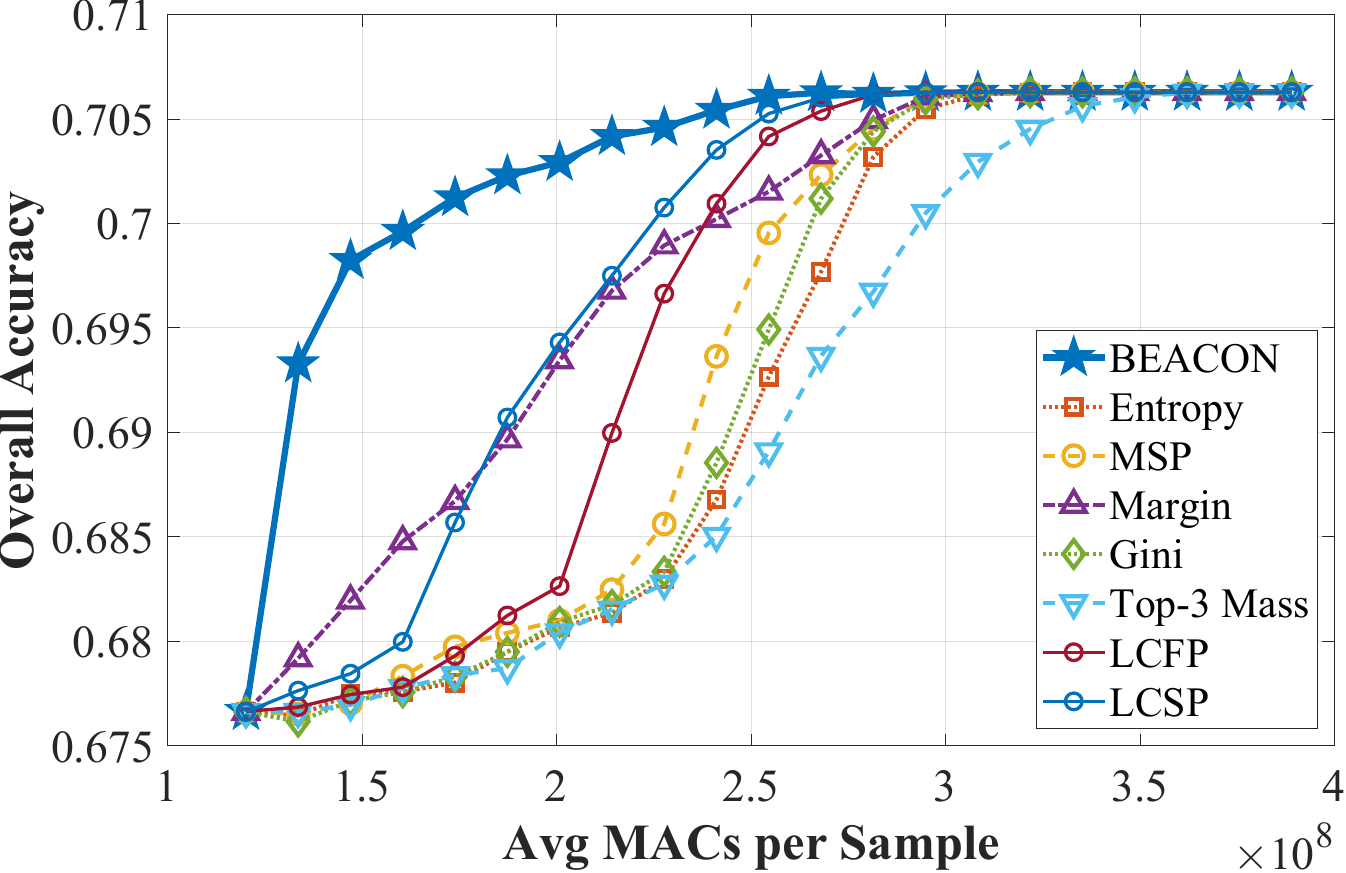}
        \label{fig:tradeoff_eel3}
    }
    \caption{Accuracy-computation trade-offs under varying thresholds for three AMC EE models.}
    \label{fig:cost_acc_tradeoff}
\end{figure*}
We next examine the  trade-off between average computational cost, measured in average MACs per sample, and overall classification accuracy. Fig.~\ref{fig:cost_acc_tradeoff} shows the results obtained by sweeping the EE threshold.  As the percentile threshold decreases from 100\% to 0\%, more samples reach the FE branch, resulting in increased computational cost but improved overall accuracy.
Note that for \textit{BEACON} and two learning-based methods, the computational costs are explicitly included in the reported  MACs.
For confidence-based baselines, the cost of computing the scoring function $\mathcal{S}(\mathbf{p}_e)$ is omitted as it is negligible compared to the predictor inference cost.

Across all three EE configurations (\textit{EE-RS1}, \textit{EE-RS2}, and \textit{EE-RS3}), the proposed \textit{BEACON} consistently outperforms  all the baselines.
With the same average MACs per sample, it achieves higher overall accuracy by more effectively identifying samples that truly benefit from deeper inference.
Conversely, under the same overall accuracy,  \textit{BEACON} requires the lowest average computational cost, demonstrating its superior efficiency in selectively invoking deeper inference only when necessary. 
\begin{table}[t]
\centering
\footnotesize
\setlength{\tabcolsep}{2pt}
\renewcommand{\arraystretch}{1.05}
\caption{Maximum achievable overall accuracy under hardware computation budgets (Avg. MACs per sample $< B$). All values are reported in percentage (\%).}
\label{tab:budget_acc_gap}
\begin{tabular}{lcccccccc}
\toprule
\textbf{$B$} & \textbf{\textit{BEACON}} & \textbf{Entropy} & \textbf{MSP} & \textbf{Margin} & \textbf{Gini} & \textbf{Top-3} & \textbf{LCFP} & \textbf{LCSP} \\
\midrule
\multicolumn{9}{l}{\textbf{\textit{EE-RS1}}}\\
$1.0\!\times\!10^{8}$ & \textbf{44.70} & 30.07 & 36.73 & 39.37 & 30.34 & 33.78 & 42.36 & 43.40 \\
$1.5\!\times\!10^{8}$ & \textbf{54.93} & 33.05 & 43.78 & 48.86 & 33.36 & 37.85 & 46.91 & 50.73 \\
$2.0\!\times\!10^{8}$ & \textbf{60.34} & 36.29 & 48.29 & 55.07 & 36.50 & 41.14 & 50.93 & 55.49 \\
$2.5\!\times\!10^{8}$ & \textbf{66.13} & 44.82 & 55.22 & 62.43 & 45.06 & 47.66 & 57.38 & 61.72\\
\midrule
\multicolumn{9}{l}{\textbf{\textit{EE-RS2}}}\\
$1.0\!\times\!10^{8}$ & \textbf{61.58} & 54.56 & 57.45 & 57.07 & 54.59 & 54.71 & 58.14 & 57.70 \\
$1.5\!\times\!10^{8}$ & \textbf{66.20} & 55.86 & 56.18 & 61.05 & 55.85 & 55.96 & 59.96 & 60.41 \\
$2.0\!\times\!10^{8}$ & \textbf{68.54} & 57.72 & 58.64 & 66.00 & 57.82 & 57.81 & 63.10 & 63.64 \\
$2.5\!\times\!10^{8}$ & \textbf{69.75} & 62.09 & 66.06 & 68.49 & 62.76 & 62.14 & 67.74 & 68.16 \\
\midrule
\multicolumn{9}{l}{\textbf{\textit{EE-RS3}}}\\
$1.5\!\times\!10^{8}$ & \textbf{69.82} & 67.75 & 67.70 & 68.20 & 67.71 & 67.70 & 67.75 & 67.85 \\
$2.0\!\times\!10^{8}$ & \textbf{70.23} & 67.95 & 68.04 & 68.97 & 67.95 & 67.87 & 68.12 & 69.07 \\
$2.5\!\times\!10^{8}$ & \textbf{70.54} & 68.68 & 69.36 & 70.02 & 68.85 & 68.50 & 70.10 & 70.35 \\
\bottomrule
\end{tabular}
\end{table}

To quantify performance under practical deployment constraints, Table~\ref{tab:budget_acc_gap} summarizes the maximum achievable overall accuracy under different computational budget constraints.
For each budget $B$, we consider all operating points on the trade-off curve and report the highest overall accuracy that satisfies $\text{Avg. MACs} < B$.
As shown in the table, \textit{BEACON} consistently achieves the highest accuracy across all budgets and EE configurations.
The performance gain is more pronounced for shallower EE settings, as a higher proportion of samples belongs to the recoverable-error case (Case $\mathcal{C}_{01}$) and therefore provides more opportunity for benefit-aware sample selection. Compared with the widely used entropy-based criterion, \textit{BEACON} achieves up to a \textbf{24\%} overall accuracy improvement under the same computational constraint, observed on \textit{EE-RS1} with a budget of $2.0 \times 10^{8}$ MACs. 
This highlights the advantage of explicitly modeling recoverability  benefits  under realistic hardware limitations.

\begin{table*}[t]
\centering
\caption{Minimum average MACs required to satisfy target overall accuracy constraints. Values in parentheses represent the cost ratio relative to the proposed method (\textit{BEACON}).}
\label{tab:acc_req_macs}
\scriptsize
\resizebox{\textwidth}{!}{
\begin{tabular}{llcccccccc}
\toprule
\textbf{Model} & \textbf{Acc. req.} & \textbf{\textit{BEACON}} & \textbf{Entropy} & \textbf{MSP} & \textbf{Margin} & \textbf{Gini} & \textbf{Top-3 mass} & \textbf{LCFP} & \textbf{LCSP}\\
\midrule

\multirow{3}{*}{{\textit{EE-RS1}}}
& $\ge 0.40$ & \textbf{74.63M} (1.00) & 222.54M (2.98) & 130.09M (1.74) & 111.61M (1.50) & 222.54M (2.98) & 185.56M (2.49) & 74.63M (1.00) & 93.12M (1.25) \\
& $\ge 0.50$ & \textbf{130.10M} (1.00) & 278.01M (2.14) & 204.05M (1.57) & 167.07M (1.28) & 278.01M (2.14) & 259.52M (1.99) & 185.56M (1.43) & 148.59M (1.14)\\
& $\ge 0.60$ & \textbf{185.56M} (1.00) & 314.98M (1.70) & 278.01M (1.50) & 222.54M (1.20) & 314.98M (1.70) & 314.98M (1.70) & 278.01M (1.50) & 222.54M (1.20)\\
\midrule

\multirow{3}{*}{{\textit{EE-RS2}}}
& $\ge 0.60$  & \textbf{86.49M} (1.00) & 220.91M (2.55) & 204.11M (2.36) & 136.90M (1.58) & 220.91M (2.55) & 220.91M (2.55) & 153.70M (1.78) & 136.90M (1.58) \\
& $\ge 0.625$ & \textbf{103.29M} (1.00) & 254.52M (2.46) & 220.91M (2.14) & 153.70M (1.49) & 237.71M (2.30) & 254.52M (2.46) & 187.31M (1.81) & 170.50M (1.65)\\
& $\ge 0.65$  & \textbf{120.10M} (1.00) & 271.32M (2.26) & 237.71M (1.98) & 187.30M (1.56) & 254.52M (2.12) & 271.32M (2.26) & 220.91M (1.84) & 220.91M (1.84)\\
\midrule

\multirow{3}{*}{{\textit{EE-RS3}}}
& $\ge 0.690$ & \textbf{133.60M} (1.00) & 254.55M (1.91) & 241.11M (1.80) & 200.79M (1.50) & 254.55M (1.91) & 267.99M (2.01) & 227.67M (1.70) & 187.35M (1.40)\\
& $\ge 0.695$ & \textbf{147.04M} (1.00) & 267.99M (1.82) & 254.55M (1.73) & 214.23M (1.46) & 267.99M (1.82) & 281.43M (1.91) & 227.67M (1.55) & 214.23M (1.46)\\
& $\ge 0.700$ & \textbf{173.91M} (1.00) & 281.43M (1.62) & 267.99M (1.54) & 241.11M (1.39) & 267.99M (1.54) & 294.87M (1.70) & 241.11M (1.39) & 227.67M (1.31)\\
\bottomrule
\end{tabular}
}
\end{table*}

Table~V summarizes the minimum average computational cost required to satisfy  target overall accuracy constraints.  
For each accuracy requirement, we report the minimum average MACs required among all percentile-based threshold operating points.
Values in parentheses denote the ratio of baseline MACs to our \textit{BEACON}'s MACs. A larger ratio indicates higher computational cost compared to \textit{BEACON}. The results show that \textit{BEACON} consistently requires the lowest average computational cost to meet a given accuracy requirement across all EE configurations.
Specifically, in the \textit{EE-RS1} configuration, the baseline methods require up to \textbf{2.98$\times$} computation cost compared to \textit{BEACON} to meet the same accuracy requirement. These results confirm that explicitly modeling recoverability, rather than just uncertainty, leads to substantially more efficient resource allocation on IoT devices.

\vspace{-2mm}
\subsection{Error Recoverability under Fixed FE Invocation Rates}

\begin{figure*}[t]
    \centering

    \subfloat[\footnotesize \textit{EE-RS1}]{
        \includegraphics[width=0.31\linewidth]{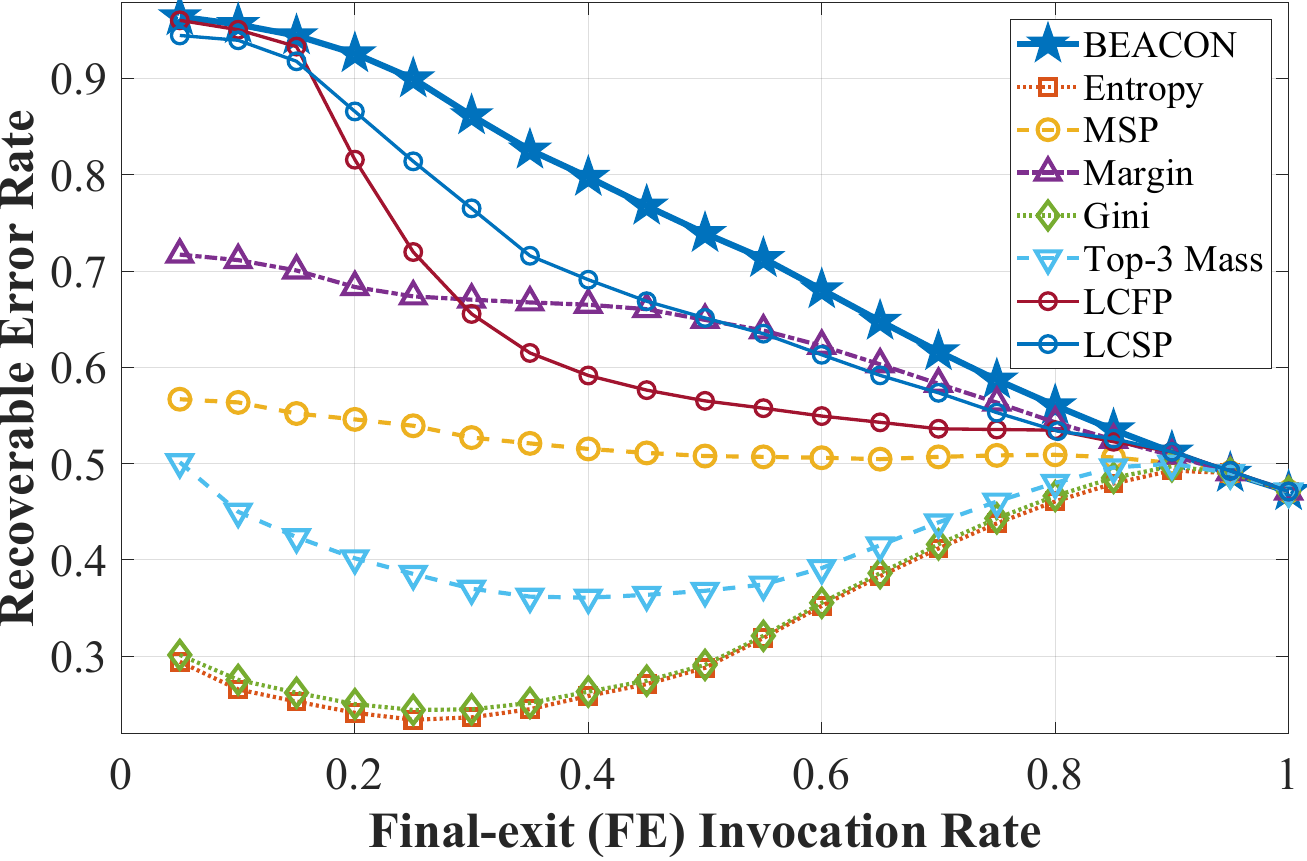}
        \label{fig:invocation_rate_eel1}
    }
    \hfill
    \subfloat[\footnotesize \textit{EE-RS2}]{
        \includegraphics[width=0.31\linewidth]{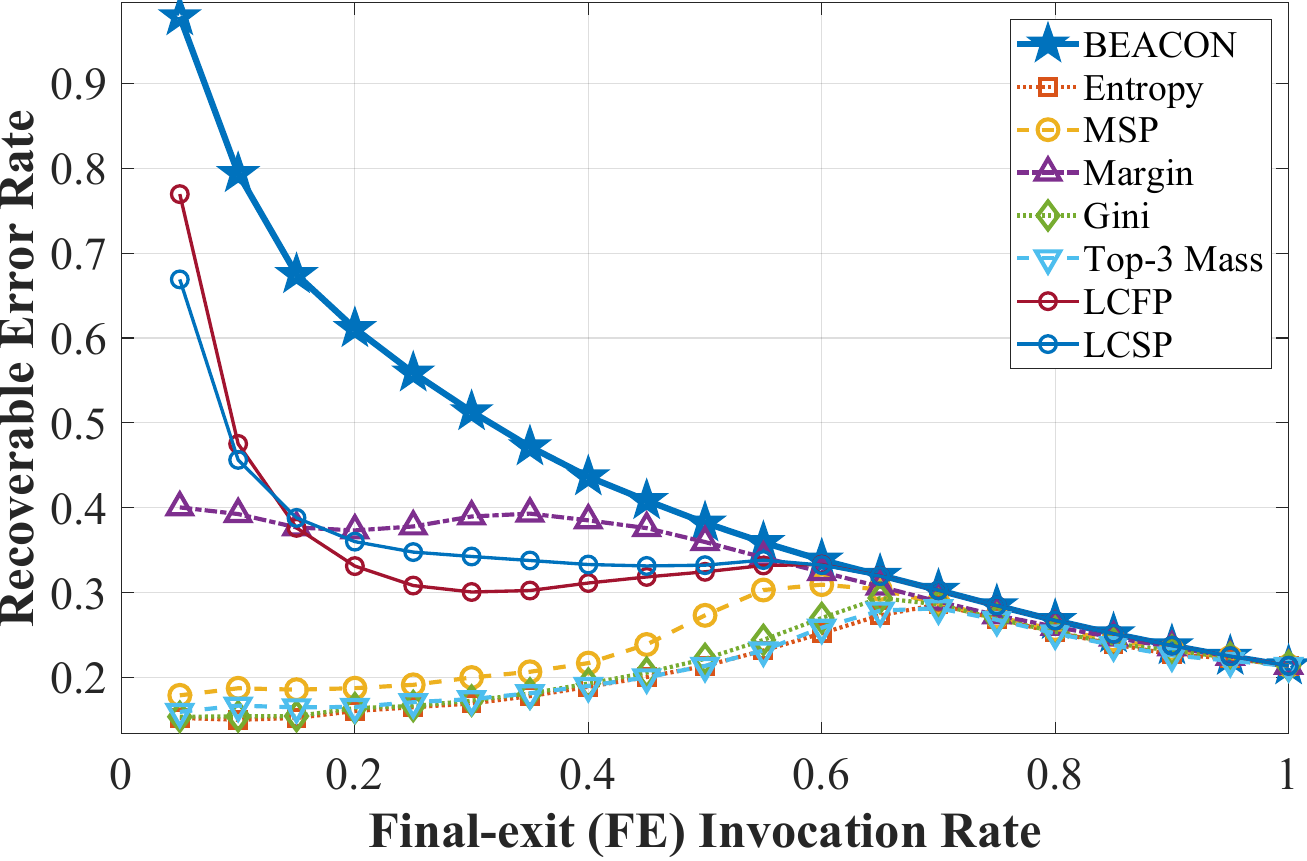}
        \label{fig:invocation_rate_eel2}
    }
    \hfill
    \subfloat[\footnotesize \textit{EE-RS3}]{
        \includegraphics[width=0.31\linewidth]{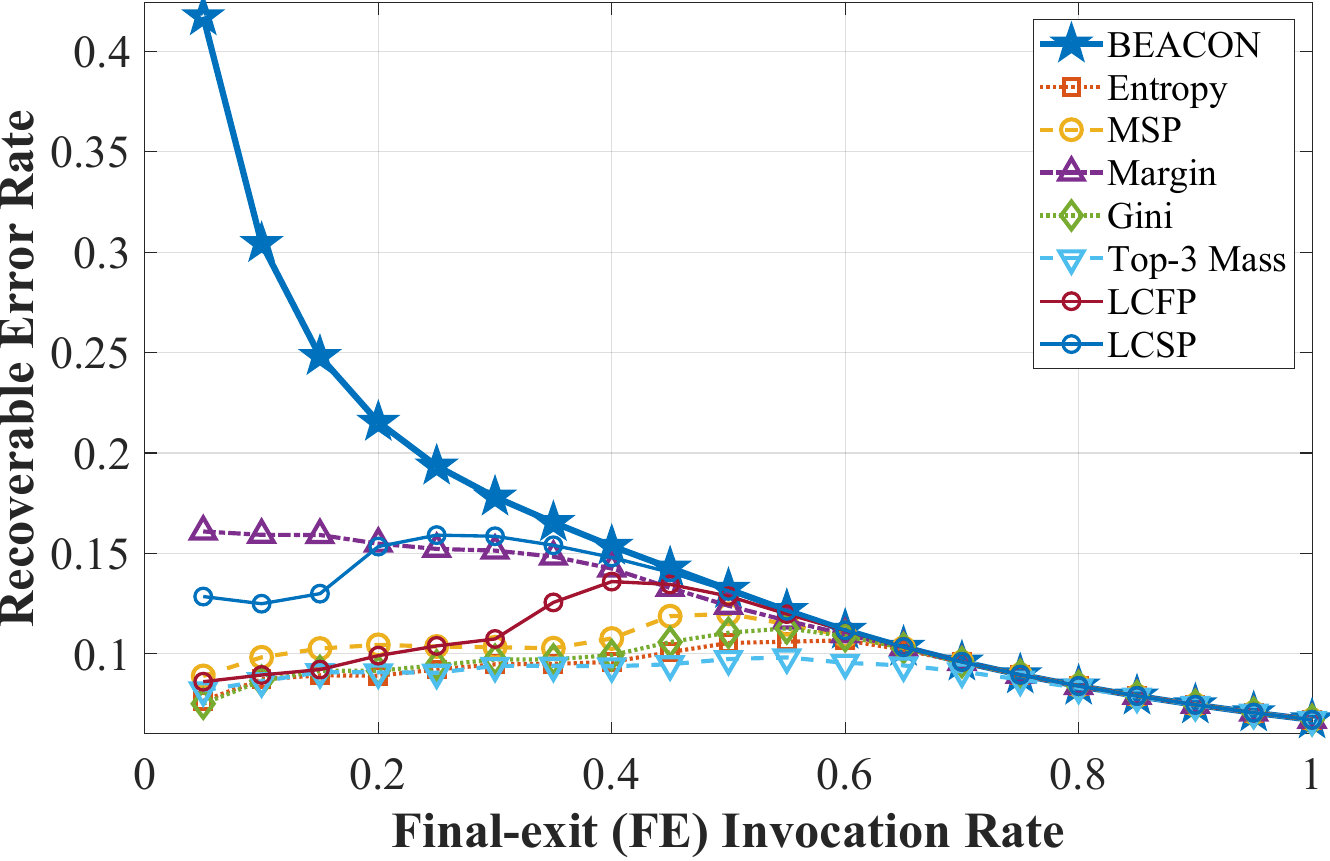}
        \label{fig:invocation_rate_eel3}
    }

    \caption{FE invocation rate versus recoverable-error rate among forwarded samples for three EE configurations.}
    \label{fig:invocation_rate}
\end{figure*}

To further analyze how effectively each criterion utilizes FE invocations, we examine the relationship between the FE invocation rate (fraction of samples forwarded to FE) and the recoverable-error rate among those forwarded samples. Specifically, we sweep the EE threshold such that the FE invocation rate increases from 5\% to 100\% in steps of 5\%.
When the invocation rate reaches 100\%, all criteria converge, as all samples are processed by the FE branch.

As shown in Fig.~\ref{fig:invocation_rate}, across all three EE configurations, \textit{BEACON} consistently achieves the highest recoverable-error rate under the same FE invocation rate.
This indicates that given a fixed budget on FE invocations, our method is  more effective at selecting samples that truly benefit from deeper inference. Since forwarding non-recoverable samples to the FE branch wastes computation resources without accuracy gain, \textit{BEACON}’s explicit focus on recoverability maximizes the utility of each invocation.
In contrast, the baseline criteria are not designed to  model recoverability, resulting in substantially lower recoverable-error rates under the same invocation constraints.

Moreover, the recoverable-error rate of \textit{BEACON}  decreases monotonically as invocation rate increases.
This behavior is consistent with the underlying benefit structure: with higher EE thresholds (i.e., lower FE invocation rates), the system becomes more selective, routing only high-probability recoverable errors to the FE branch.
In contrast,  all baselines show erratic, non-monotonic trends, highlighting the fundamental mismatch between confidence or consistency scores and the actual benefit of deeper inference.


\subsection{SNR-Dependent Computation-Accuracy Trade-offs}
\begin{figure*}[t]
    \centering
    \subfloat[\footnotesize High SNR:  (10--20~dB)]{
        \includegraphics[width=0.22\linewidth]{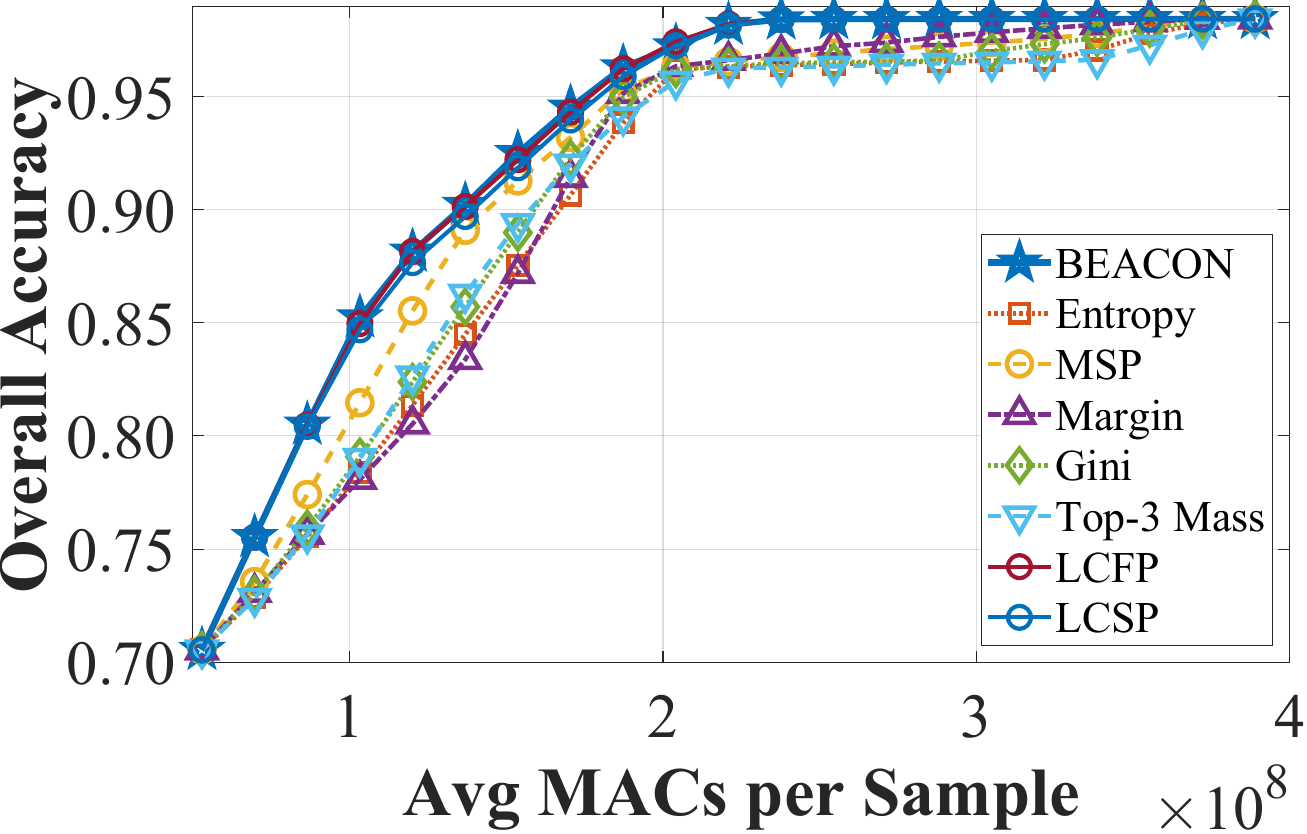}
        \label{fig:snr_high}
    }
    \hfill
    \subfloat[\footnotesize Medium SNR: (0--10~dB)]{
        \includegraphics[width=0.22\linewidth]{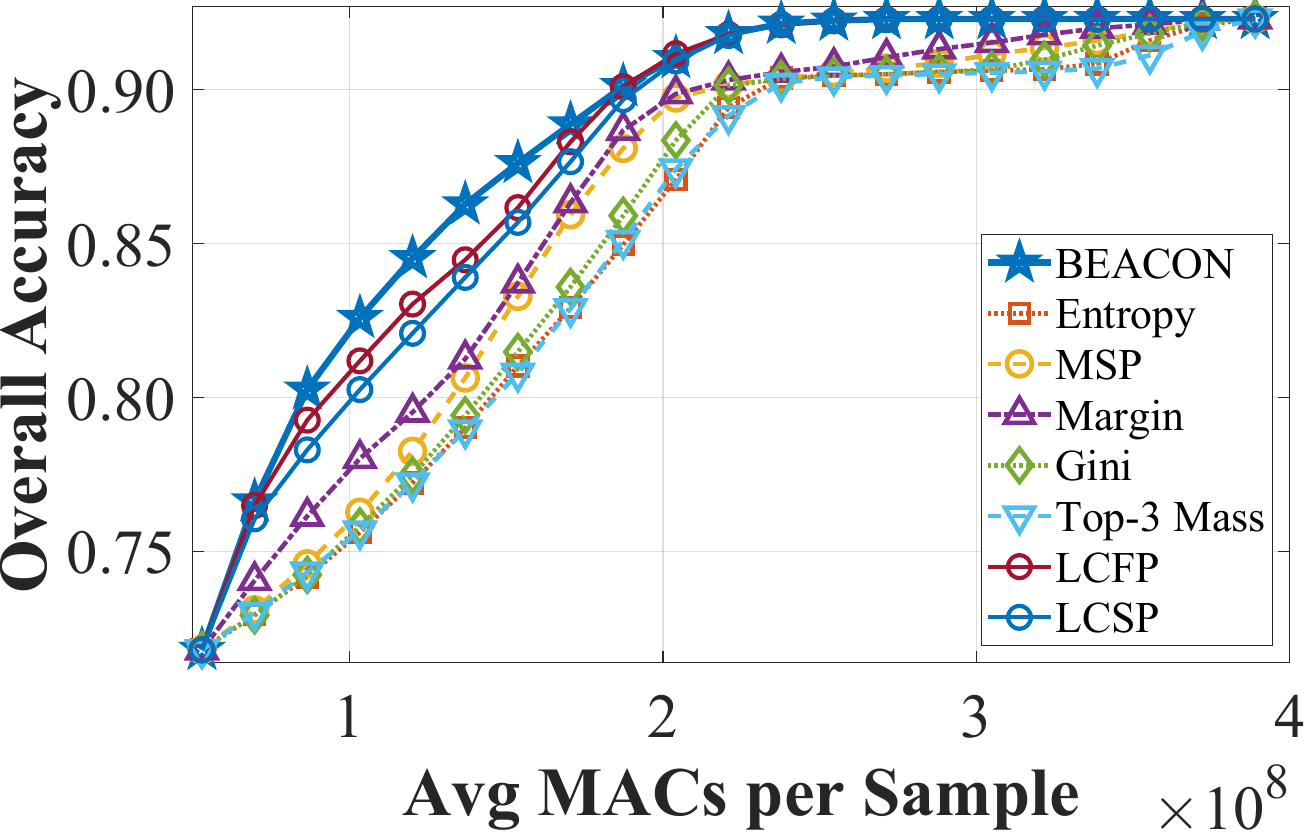}
        \label{fig:snr_medium_high}
    }
    \hfill
    \subfloat[\footnotesize Low SNR: (-10--0~dB)]{
        \includegraphics[width=0.22\linewidth]{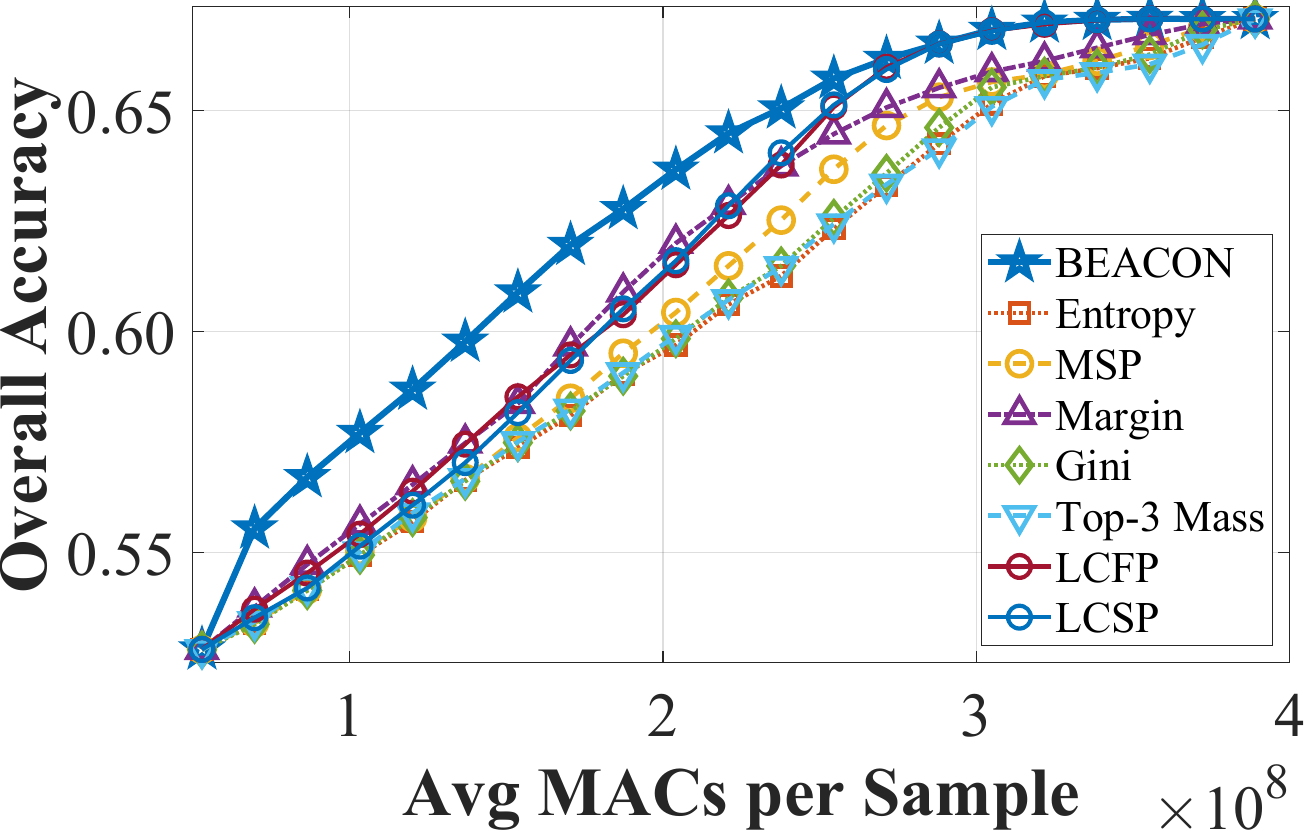}
        \label{fig:snr_medium_low}
    }
    \hfill
    \subfloat[\footnotesize Very Low SNR: (-20--10~dB)]{
        \includegraphics[width=0.22\linewidth]{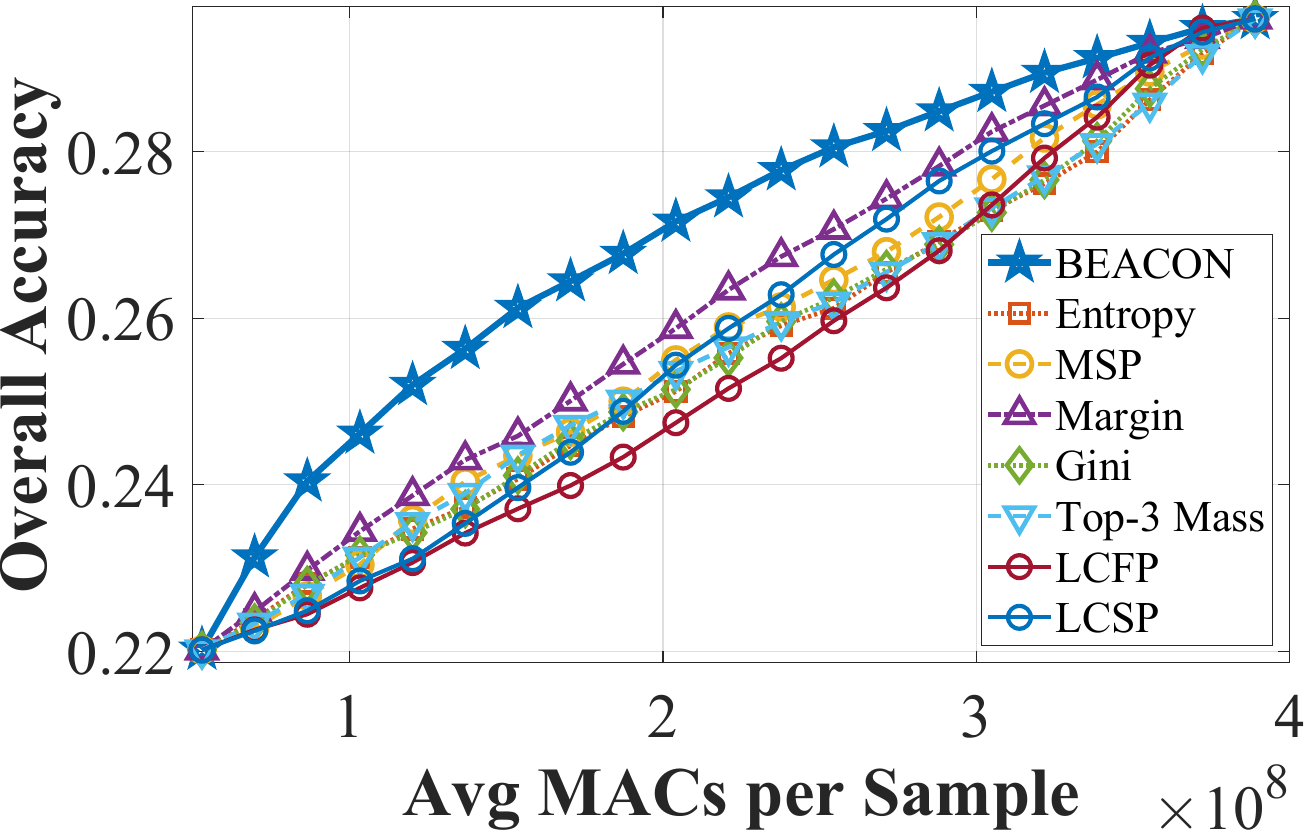}
        \label{fig:snr_low}
    }
    \caption{SNR-dependent computation--accuracy trade-offs for the \textit{EE-RS2} AMC EE model.}
    \label{fig:snr_tradeoff}
\end{figure*}

To evaluate  robustness under varying channel conditions, we  conduct an SNR-dependent analysis on the \textit{EE-RS2} model. The results for \textit{EE-RS1} and \textit{EE-RS3} exhibit similar trends and are  omitted for brevity.
We group the test samples into four SNR ranges:  high SNR (10-20~dB), medium SNR (0-10~dB), low SNR (-10-0~dB), and very low SNR (-20 to -10~dB).

Figure~\ref{fig:snr_tradeoff} shows that \textit{BEACON} consistently achieves better  computation-accuracy trade-offs than confidence-based and learning-based baselines across all SNR regimes.
For a given average computational cost, it achieves higher overall classification accuracy, and conversely, it requires less computation to reach the same accuracy level.
At high SNR, where most samples are relatively easy to classify, all methods approach high accuracy with sufficient computation.
Nevertheless, \textit{BEACON} still exhibits a clear advantage in the low-computation regime by enabling earlier termination with minimal accuracy degradation.
With medium and low SNR, where recoverable errors are more prevalent, the performance gap becomes more pronounced.
In the very low SNR regime, despite the overall low accuracy, \textit{BEACON} consistently maintains the best accuracy-computation trade-off.
These results confirm that \textit{BEACON} remains effective and robust under diverse channel conditions, making it suitable for dynamic wireless environments.

\subsection{Realiability of \textit{BEACON}}
\begin{table}[t]
\centering
\caption{Calibration of \textit{BEACON} for different EE models.}
\label{tab:predictor_calibration}

\vspace{1mm}
\resizebox{\linewidth}{!}{%
\begin{tabular}{c c c c}
\toprule
\textbf{Model} &
\textbf{Avg. Predicted Prob.} &
\textbf{True Recoverable Ratio} &
\textbf{Abs. Gap} \\
\midrule
\textit{EE-RS1} & 0.4802 & 0.4710 & 0.0092 \\
\textit{EE-RS2} & 0.2226 & 0.2139 & 0.0087 \\
\textit{EE-RS3} & 0.0711 & 0.0672 & 0.0039 \\
\bottomrule
\end{tabular}}
\end{table}

We assess the reliability of \textit{BEACON} by comparing its predicted recoverable probabilities with 
the empirically observed recoverable-error ratios on the test set. 
As shown in Table~\ref{tab:predictor_calibration},
across all three EE configurations, the predicted recoverable
probabilities closely match the actual recoverable-error ratios, with 
minor absolute gaps (below 1\%).
This high degree of calibration  demonstrates that \textit{BEACON} provides a meaningful
probabilistic estimate of recoverability rather than an arbitrary
score.
As a result, the threshold used in the benefit-aware EE decision rule
admits a clear probabilistic interpretation, enabling intuitive and reliable
control of the accuracy-computation trade-off.
This calibration property further explains the superior performance of \textit{BEACON} observed in previous analyses.
\subsection{Generalizability of \textit{BEACON}}

\begin{figure*}[t]
    \centering

    \subfloat[\footnotesize \textit{Model V0}]{
        \includegraphics[width=0.31\linewidth]{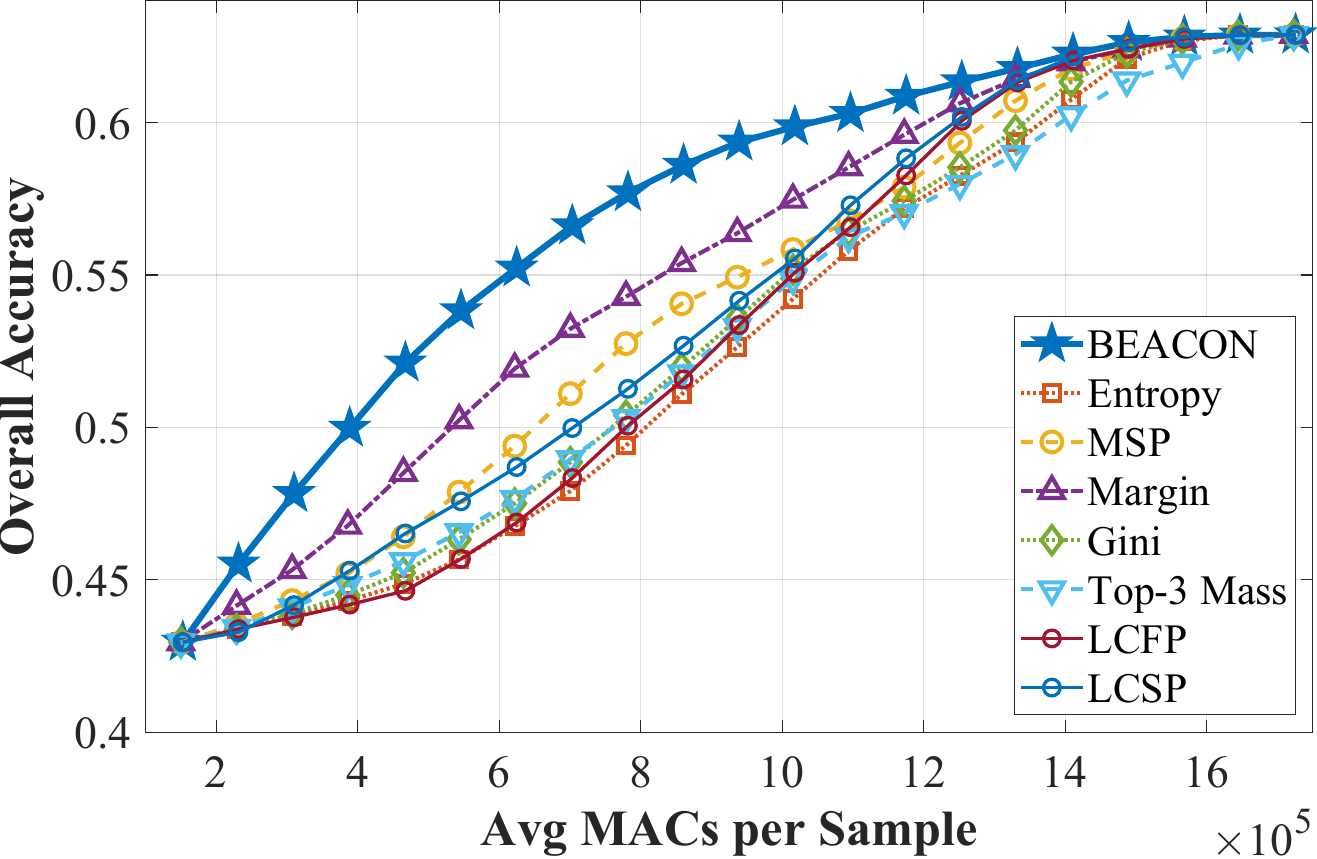}
        \label{fig:invocation_rate_eel11}
    }
    \hfill
    \subfloat[\footnotesize \textit{Model V1}]{
        \includegraphics[width=0.31\linewidth]{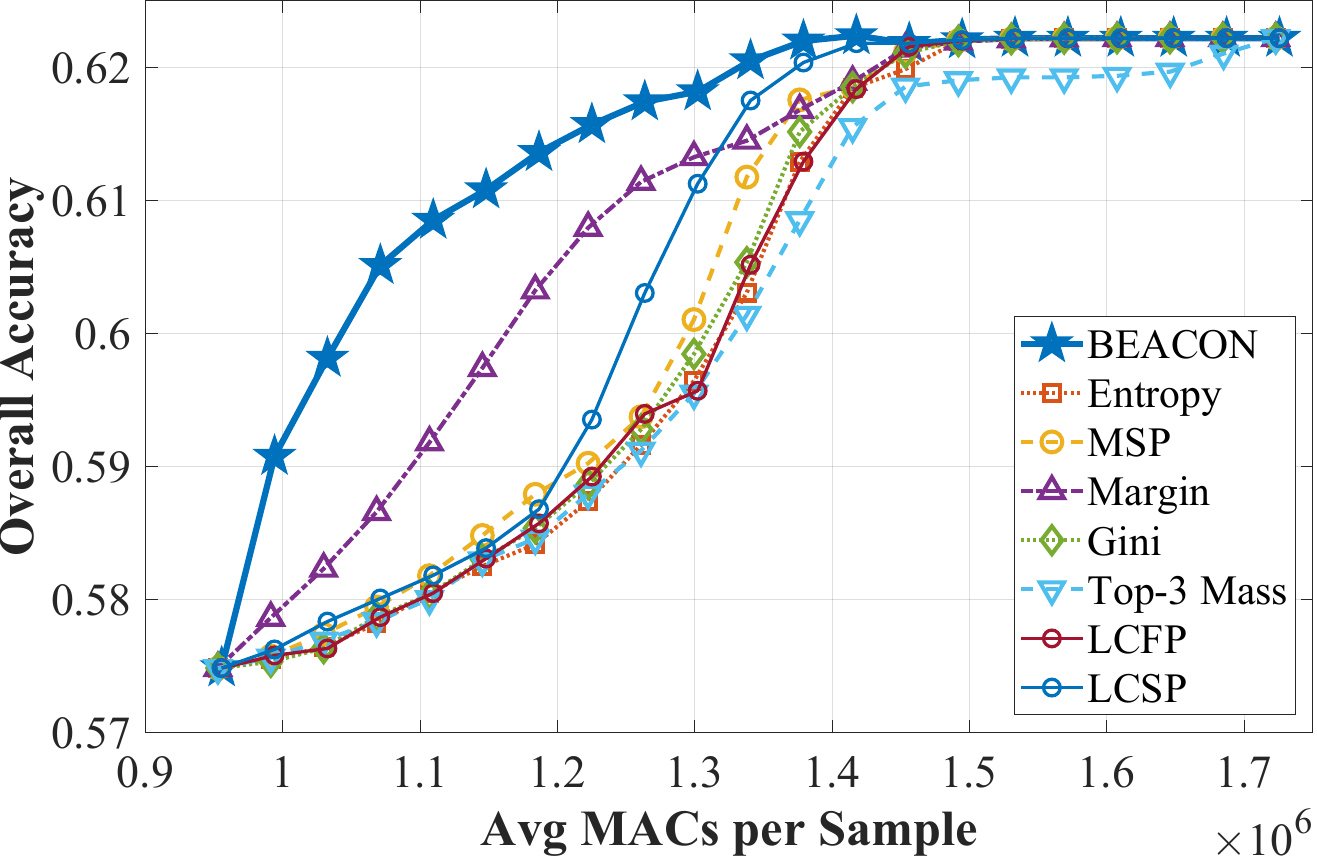}
        \label{fig:invocation_rate_eel22}
    }
    \hfill
    \subfloat[\footnotesize \textit{Model V2}]{
        \includegraphics[width=0.31\linewidth]{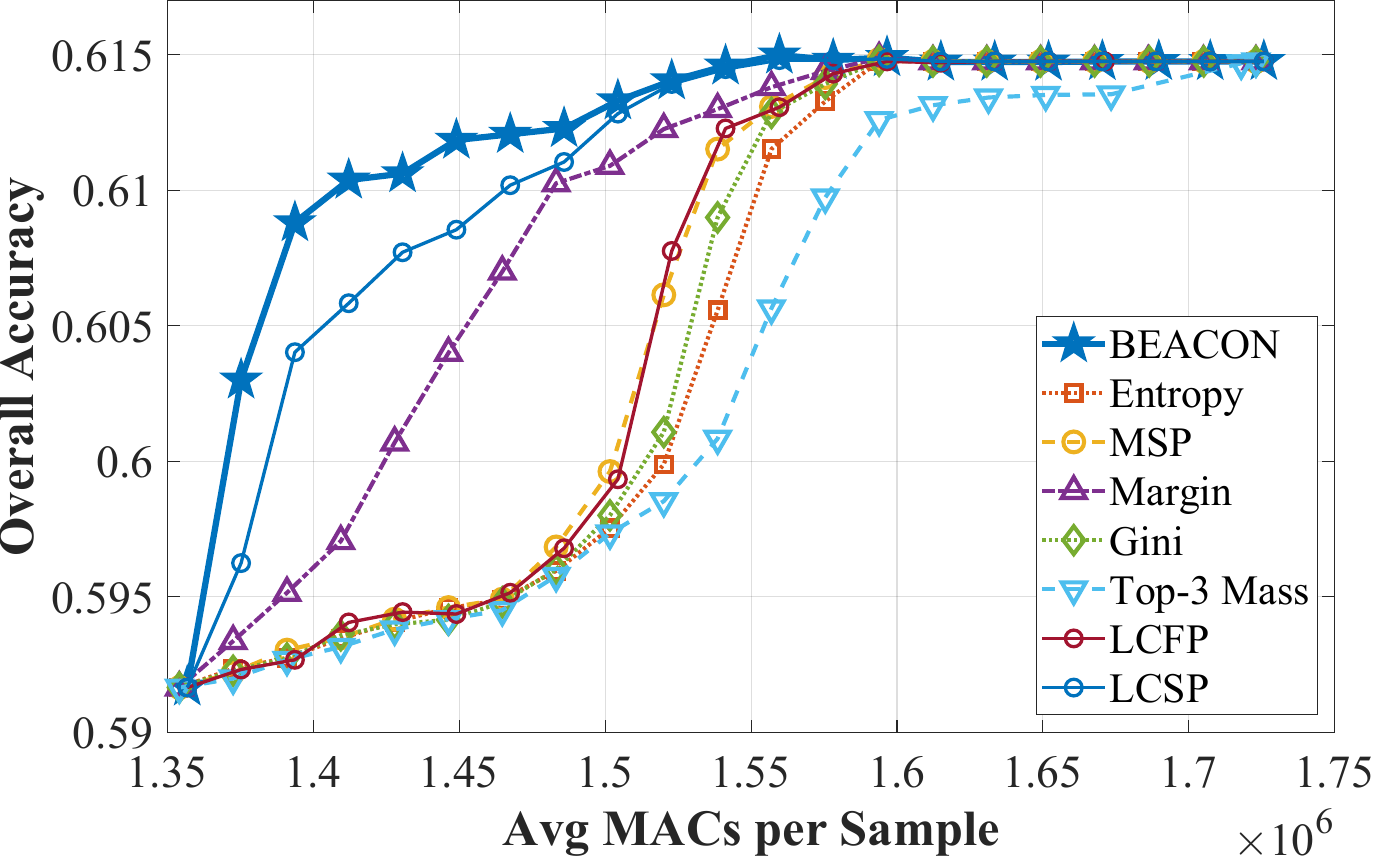}
        \label{fig:invocation_rate_eel33}
    }

    \caption{Accuracy-computation trade-offs under varying thresholds for the three EE architectures (V0, V1, and V2), demonstrating the generalizability of \textit{BEACON} across different EE configurations.}
    \label{fig:invocation_rate1}
    \vspace{-2mm}
\end{figure*}

To further evaluate the generalizability of \textit{BEACON}, we additionally conduct experiments on the three EE architectures (\textit{V0}, \textit{V1}, and \textit{V2}) proposed in \cite{elsayed2023earlyexit}, one of the first works introducing EE strategies for AMC. These models are built upon the same sequential CNN backbone architecture, consisting of stacked 1D convolutional blocks followed by fully connected layers, while adopting different EE insertion positions. As a result, they exhibit different accuracy-computation trade-offs and different gaps between EE and FE predictions.

Figure~\ref{fig:invocation_rate1} shows the trade-off curves between average computational cost, measured in average MACs per sample, and overall classification accuracy for all three architectures. Similar to Fig.~\ref{fig:cost_acc_tradeoff}, the curves are obtained by sweeping the EE thresholds. As the percentile threshold decreases from 100\% to 0\%, more samples are forwarded to the FE branch, resulting in increased computational cost but improved overall accuracy.

Across all model configurations, \textit{BEACON} consistently achieves the best accuracy-computation trade-off and outperforms both confidence-based and learning-based EE criteria. In particular, \textit{BEACON} achieves clear performance advantages in the low- and medium-computation regions, demonstrating that the proposed benefit-aware criterion generalizes effectively across different EE configurations.

Since the three models use the same backbone while placing the EE branch at different network depths, their EE accuracies differ significantly. Specifically, V0 adopts the shallowest EE position and therefore exhibits the lowest EE accuracy and the largest EE--FE performance gap. In contrast, V2 places the EE branch deeper in the network, resulting in higher EE accuracy and a smaller EE--FE gap, while V1 provides an intermediate configuration. The consistent improvements achieved by \textit{BEACON} across all three settings further demonstrate that the proposed recoverability prediction mechanism is not tied to a specific EE position or model configuration.

\section{Discussions}

In this section, we discuss some important aspects and potential extensions of \textit{BEACON}, including the use of Top-$k$ probability representations and the relationship between \textit{BEACON} and existing width-adaptive and pruning-based methods.

\subsection{TOP-$k$ Probability for Large Output Spaces}

Using only Top-$k$ probabilities as the predictor input can further reduce the complexity of LBAP. In our implementation, replacing the full 10-dimensional output distribution with a Top-3 representation reduces the predictor MACs and parameter count by approximately 16\%. However, as shown in Table~\ref{tab:overhead}, the overall overhead of LBAP is already extremely small compared to the backbone network, accounting for only $0.0007\%$ of the total MACs and $0.0252\%$ of the parameters. Therefore, the additional savings brought by Top-$k$ compression are practically negligible in the AMC setting considered in this work, where the number of modulation classes is inherently small. Nevertheless, Top-$k$-based representations may become considerably more valuable in future large-scale classification tasks with much larger output spaces, such as large language models (LLMs) with vocabularies containing tens of thousands of tokens. In such scenarios, restricting the predictor input to Top-$k$ outputs could significantly reduce memory access, computational complexity, and predictor size. However, reducing the input dimension may also decrease the benefit prediction accuracy due to the loss of information contained in lower-probability classes. Therefore, an important future research direction is to investigate the trade-off between predictor performance and the choice of $k$.

\subsection{Relationship with Existing Adaptive Inference Methods}

As an adaptive inference method, 
\textit{BEACON} belongs to the  category of depth-adaptive inference, where samples dynamically determine whether additional layers should be executed. In contrast, model compression and width-adaptive approaches, such as network pruning \cite{han2015learning,liu2017learning} and slimmable networks \cite{yu2019slimmable,yu2019universally}, primarily improve efficiency by reducing or dynamically adjusting model width and computational capacity. Although both paradigms aim to improve the accuracy-efficiency trade-off, their adaptation mechanisms are fundamentally different.
Specifically, \textit{BEACON} focuses on predicting the \emph{benefit of continuing inference}, i.e., whether forwarding a sample to deeper layers is likely to correct the EE prediction. Existing pruning and width-adaptive methods typically allocate computational resources according to learned gating policies, confidence-related signals, predefined subnet configurations, or hardware/resource constraints. However, these approaches generally do not explicitly model the recoverability of EE errors, which is the primary objective of \textit{BEACON}. Therefore, the proposed benefit-aware criterion is complementary to width-adaptive techniques.

Moreover, \textit{BEACON} is largely architecture-agnostic and can potentially be integrated with pruning-based or slimmable architectures to jointly adapt both inference depth and network width. For example, future systems may simultaneously determine (i) whether deeper inference is necessary and (ii) how much model capacity should be allocated for the current sample. We believe such joint depth-width adaptive inference is a promising future research direction.
\section{Conclusions}

This paper studied a fundamental question in EE AMC inference on resource-constrained IoT devices: \emph{when is deeper inference truly beneficial?} Through systematic analysis, we showed that conventional confidence-based EE criteria are unreliable indicators of the actual \emph{benefit} of invoking deeper layers. To bridge this gap, we proposed \textit{BEACON}, a novel benefit-aware EE framework. By explicitly predicting  \emph{recoverable errors}, \textit{BEACON} introduces a principled criterion that quantifies expected accuracy gain of deeper inference. We further implemented this criterion by developing the LBAP, which utilizes the full EE probability vector to retain class-competition structure that is lost in scalar confidence scores. Experiments on three ResNet-18-based AMC EE models demonstrated consistently superior computation-accuracy trade-offs compared with representative baselines.

Beyond empirical gains, this work introduces a new design principle for EE networks: \textit{shifting the objective from generic confidence estimation to explicit benefit prediction}. Our results show that modeling error correction potential is essential for efficient inference. By identifying samples that can truly benefit from deeper processing, our method enables more principled and efficient computation allocation, particularly under strict resource budgets and varying SNR conditions.
Moving forward, we plan to extend this benefit-aware framework to multi-exit architectures and explore its application to broader physical-layer intelligence tasks, such as signal detection and interference identification, to support the evolving requirements of autonomous IoT ecosystems.

\bibliographystyle{IEEEtran}
\bibliography{ref}

\vfill

\end{document}